\documentclass[a4paper,11pt]{article}
\pdfoutput=1 

\usepackage{jcappub} 
\usepackage[T1]{fontenc}
\usepackage[utf8]{inputenc}

\usepackage{amsmath}
\usepackage{bm}
\usepackage{txfonts}

\usepackage[protrusion=true,expansion=true]{microtype}
\usepackage{graphicx}
\usepackage{upgreek}

\usepackage{array,multirow,makecell}
\usepackage{booktabs}

\usepackage{xcolor}
\usepackage{hyperref}

\usepackage[mathlines]{lineno}

\newcommand{\mathbfcal}[1]{\bm{\mathcal{#1}}}
\def\rbm#1{\xrbm#1\relax^\relax\valign}
\def\xrbm#1^#2\relax#3\valign{%
\mathbf{#1}\ifx\valign#2\valign\else^{\mathbf{#2}}\fi}

\title{\boldmath Reconstruction of cosmic-ray direction and energy in radio arrays using deep ensemble graph neural networks}

\author[a,b]{Arsène Ferrière,\note{Corresponding author.}}
\author[a]{Aurélien Benoit-Lévy,}
\author[b,c]{Olivier Martineau-Huynh}
\author[d]{Matías Tueros}

\affiliation[a]{Universit\'{e} Paris-Saclay, CEA, List\\ F-91120 Palaiseau, France}
\affiliation[b]{Sorbonne Université, CNRS, Laboratoire de Physique Nucléaire et des Hautes Énergies (LPNHE)\\ 4 Pl. Jussieu, 75005 Paris, France}
\affiliation[c]{Sorbonne Universit\'{e}, UPMC Univ. Paris 6 et CNRS, UMR 7095, Institut d'Astrophysique de Paris\\ 98 bis bd Arago, 75014 Paris, France}
\affiliation[d]{Instituto de Física La Plata, CONICET - UNLP\\ Boulevard 120 y 63 (1900), La Plata - Buenos Aires, Argentina}

\emailAdd{arsene.ferriere@proton.me}
\emailAdd{aurelien.benoit-levy@cea.fr}
\emailAdd{martineau@lpnhe.in2p3.fr}
\emailAdd{mjtueros@gmail.com}

\abstract{
Using advanced machine learning techniques, we developed a method to reconstruct the arrival direction and energy of ultra-high-energy cosmic rays from the voltage traces they induce on ground-based radio detector arrays.

In our approach, triggered antennas are represented as a graph structure, which serves as input for a graph neural network (GNN). By incorporating physical knowledge into both the GNN architecture and the input data, we improve the precision and reduce the required size of the training set with respect to a fully data-driven approach. This method achieves an angular resolution of 0.092° and an electromagnetic energy reconstruction resolution of 16.4\% on simulated data with realistic noise conditions.

We also employ uncertainty estimation methods to enhance the reliability of our predictions, quantifying the confidence of the GNN’s outputs and providing confidence intervals for both direction and energy reconstruction.
Finally, we investigate strategies to verify the model’s consistency and robustness under real life variations, with the goal of identifying scenarios in which predictions remain reliable despite domain shifts between simulation and reality.
}
    
\keywords{ultra-high-energy cosmic rays, radio detection, direction reconstruction, graph neural network, energy reconstruction, deep learning}
\arxivnumber{2602.23321}

\begin{document}
\maketitle
\flushbottom

\section{Introduction}
When a high-energy cosmic ray enters the Earth's atmosphere, it interacts with air molecules and initiates a cascade of secondary particles. The detection of this cascade, known as an extensive air shower (EAS), forms the basis for the study of cosmic rays at energies typically above $10^{15}$\,eV. Various detection techniques are used, among which scintillator arrays, water-Cherenkov detectors, fluorescence telescopes, and radio antennas are the most common (see e.g. \cite{Coleman:2022abf} for a detailed review).

Radio antennas detect the radiation emitted by the EAS through the geomagnetic and charge-excess (or Askaryan) effects~\cite{Kahn1966}. This emission is coherent --and thus detectable-- for wavelengths larger than the meter-scale shower pancake, i.e. frequencies in the tens of MHz range. Radio antennas are inexpensive, robust and stable detectors. They are also easier to deploy than most other techniques, while the atmosphere is transparent at these frequencies. Radio detection is therefore particularly well suited for the observation of cosmic particles with low rates, which require large detection areas, and very inclined showers, which develop far away from the detector.

The Auger collaboration has thus equiped its Surface Detector with radio antennas to study ultra-high-energy cosmic rays (UHECRs)~\cite{AugerPrime_2019}, while the GRAND collaboration proposes using a network of giant radio arrays to observe ultra-high-energy particles ($\geq 10^{17}$\,eV)~\cite{GRAND_paper_2019}. In its prototype phase, whose main instrument is GRANDProto300, a 300-antenna setup currently being deployed in Gansu Province, China, GRAND also targets EAS induced by cosmic rays~\cite{GRAND_ICRC2025}.


Precise reconstruction of air-shower parameters is essential for advancing the analyses of radio data produced by these detectors. Accurate directional reconstruction will also be mandatory for neutrino astronomy. To this end, several reconstruction techniques have been developed. Classical likelihood-based methods such as plane wavefront (PWF) fitting~\cite{Ferriere2024, CORSTANJE201522} can provide good estimates of shower direction. More refined techniques, like the angular distribution function (ADF)~\cite{GUELFAND2025} or the lateral distribution function (LDF)~\cite{LDF_Auger, Guelzow_ICRC}, offer improved reconstruction of both direction and energy from the electric fields recorded by the antennas. However, these approaches can be affected by array sparsity and inhomogeneity, threshold effects in selection of detection units, and imperfect calibration, which may introduce biases at large zenith angles or low signal-to-noise ratios. They also rely on empirical models of radio signal distribution that do not fully capture the underlying physics.

In this work, we introduce a machine learning approach to estimate the direction of the primary particle and the energy of the EAS electromagnetic component, offering robust predictions with associated uncertainty estimates. Our method is hybrid, combining an EAS-physics-informed estimator with a data-driven model trained on simulations to improve upon the classical PWF estimate. 

The machine learning models employed in this work are Graph Neural Networks (GNNs), 
because they naturally handle variable-size, irregularly spaced antenna layouts and are permutation-invariant by design. They are therefore well suited to autonomous, distributed arrays where the number of triggered antennas and the shape of the footprint can vary widely. 

GNNs have already shown promising results in related experiments such as ALPAQUITA~\cite{Garcia_2025} and IceCube~\cite{Koundal_2025}, despite differences in experimental configurations, model architectures and reconstruction goal. 
Yet to our knowledge, uncertainty quantification for GNN-based reconstruction has only recently begun to be explored in astroparticle physics.
 Simulation-based inference with a GNN and a normalizing-flow posterior has for example been applied for the reconstruction of ultra-high-energy cosmic-ray directions from radio arrays, achieving calibrated uncertainty estimates in a work inspired by  ~\cite{FerriereICRC} and presented in ~\cite{Macias2025Direction}, a publication to which some of us also contributed. In the present work, we propose a different approach to uncertainty estimation, based on probabilistic regression models and deep ensembles, and perform a quantitative calibration study. This is essential for downstream scientific analyses that rely on reliable confidence regions. 

In the next section, we describe the simulation and signal-conversion pipeline used to generate training and evaluation datasets, including realistic antenna and RF-chain responses and noise modeling. In Section~\ref{section:methods}, we present the methodology and model architectures, including the physics-informed inputs and uncertainty estimation strategy. In Section~\ref{section:performances}, we evaluate performances of the method, and discuss uncertainties and robustness in Sections \ref{sec:uncertainty_estimations} and \ref{sec:robustness} respectively. We conclude in Section~\ref{section:conclusions}.

\section{Simulation and Signal Conversion}
Realistic simulations of voltage signals induced by EAS are critical to train and evaluate the reconstruction models presented in section \ref{section:methods}. We detail here the process we used to generate them. 

\subsection{Electric field generation}
We simulated 25'000 EAS using the ZHAireS Monte Carlo code~\cite{Alvarez_Muniz_2012} with zenith angles $\uptheta$ following a $-\log{(\cos{\uptheta})}$ distribution between 0° (vertical) and 90° (horizontal). Azimuth angles $\upphi$ are sampled uniformly between 0° and 360°. Energies are drawn from a logarithmic distribution between 0.4 and 4\,EeV. Proton and iron primaries are considered in equal proportions. Shower cores are randomly distributed above a GRAND-like hexagonal array~\cite{layout2024}.

For each simulated event, electric field values are computed at antenna location withing a 3° off-axis angle from the axis from the layout (see Figure~\ref{fig:graph_exemple}), generating 4\,$\mu$s-long timetraces with a 0.5\,ns step along the 3 directions X (South-North), Y (West-East), Z (Up).

\subsection{Electric field conversion to voltage}
\label{sec:voltage}
\begin{figure}[t]
\centering 
\includegraphics[width=0.44\textwidth,trim={0 0 0 1.3cm}, clip]{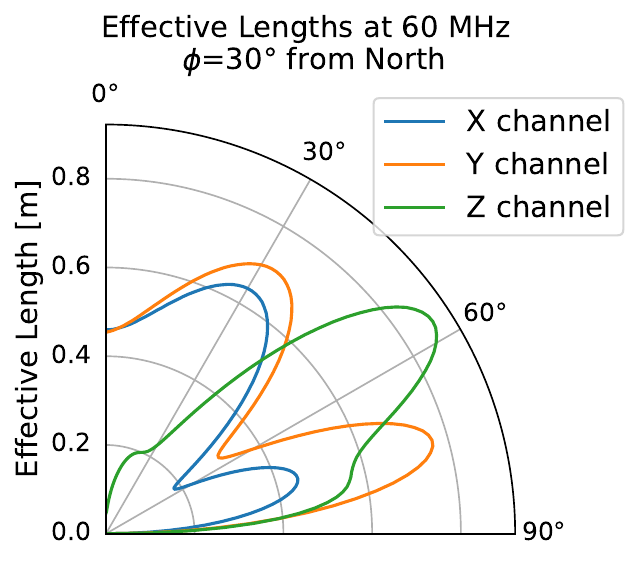} \includegraphics[width=0.55\textwidth]{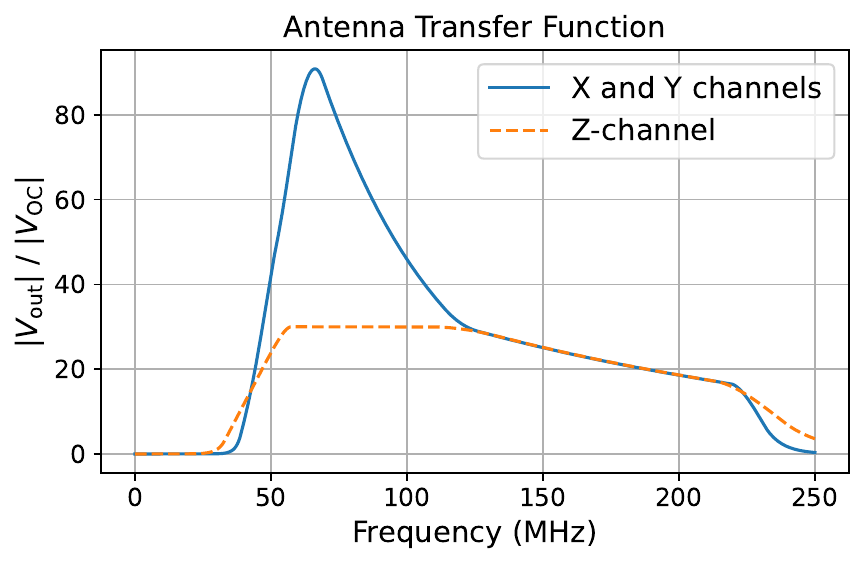} 
\caption{Left: Magnitude of the modeled RF chain transfer function for the X and Y (solid blue) and Z (dashed orange) antenna arms. The magnitude response shows frequency-dependent gain variations. Right: Effective lengths of the horizontal antennas at 60 MHz for a signal coming 30° from north in the presence of ground reflection as a function of the zenith angle.}
\label{fig:rfchain} 
\end{figure}

The electric-field time traces are then converted into voltage signals using a realistic antenna response model. To do so, we use the geometry of the so-called HorizonAntenna proposed by the GRAND collaboration~\cite{GRAND_paper_2019, GRANDHW} and use the computet effective length $\rbm{l}_{\rm eff}(\theta, \phi, \nu)$\,\cite{balanis01} from the NEC4 simulation code~\cite{Burke2004NEC}. A visualization of the effective lengths used in this study at 60\,MHz is shown in Figure~\ref{fig:rfchain}.


The open-circuit voltage $V_{{\rm oc},\, a}(\nu)$ on a given arm $a \in \{X, Y, Z\}$ is then computed in the frequency domain as:
\begin{equation}
V_{{\rm oc},\, a}(\nu) = \rbm{l}_{{\rm eff},\, a}(\theta_{\max}, \phi_{\max}, \nu) \cdot \rbm{E}(\nu),
\end{equation}
where $\rbm{E}(\nu)$ is the electric field vector at the antenna location and $(\theta_{\max}, \phi_{\max})$ corresponds to the direction of the incoming wave as seen from the antenna position, under the assumption that the radio emission originates from the shower maximum $X_{\rm max}$.

The resulting open-circuit voltage is then fed into a generic analog electronic chain to obtain the output voltage recorded by the detector. In this work, we adopt a frequency-domain transfer function inspired by the GRAND RF-chain response~\cite{GRANDLIB_2025, GRANDProtoPapers_2025}. It is approximated from the RF-chain response as a smoothed piecewise-linear function in logarithmic frequency space between 50 and 250~MHz. Its gain is diplsayed in Figure~\ref{fig:rfchain}. The transfer function is assumed to be identical for the horizontal arms (X and Y), while small variations are introduced for the vertical (Z) arm. The phase response is modeled as a linear function of frequency, corresponding to a constant time delay without distortion of the signal shape.

The effect of the electronics is represented as a diagonal matrix $\mathbfcal{R}(\nu)$ acting independently on each channel of $\rbm{V}_{\rm oc}$. The final output voltage vector is then given by:
\begin{equation}
\rbm{V}_{\rm out}(\nu) = \mathbfcal{R}(\nu)\cdot \rbm{V}_{\rm oc}(\nu).
\end{equation}
This procedure allows us to generate transient voltage pulses that are similar to those obtained with the standard GRAND simulation framework~\cite{GRANDLIB_2025}. These simulated pulses also exhibit very similar features to those measured by the GRANDProto300 array in response to cosmic-ray-induced air shower candidates (see e.g. \cite{NUTRIG}). Yet the detector response model is kept sufficiently generic so that reconstruction methods trained on these simulations are not specific to GRAND and can be adapted to other radio detection experiments.

\subsection{Noise modeling}
\label{sec:noise}
To make the simulated data suitable for training, it is essential to add realistic experimental nuisance to the simulated voltage pulses. However, here again we aim to keep these characteristics as generic as possible, so that reconstruction methods developed using this pulse library are not tied to the specific case of GRANDProto300. Therefore, the simulated signal is altered through three distinct operations.

\begin{figure}[t]
        \centering
        \includegraphics[width=0.75\textwidth]{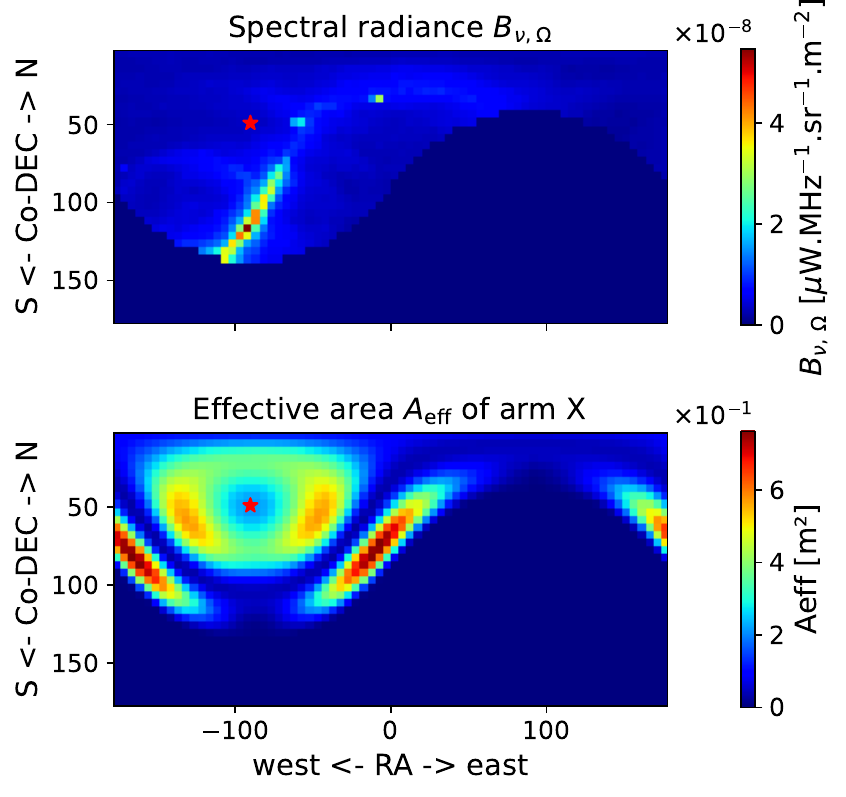}
        \caption{Spectral radiance map $B_{\nu}$ derived from the temperature map (top) and effective area map ($A_{\rm eff}$) (bottom) for arm X of the detector at 60 MHz as a function of right ascension (RA) and latitude. The red star indicates the zenith at detector's location. Both plots correspond to a local sidereal time of 18:00.}\label{fig:galatic_emission}
\end{figure}

\begin{itemize}
    \item First, we consider the Galactic noise contribution to the ambient electromagnetic background. 
    It can be treated as an incoherent stationary signal added to the simulated trace after computation through a standard treatment already exposed in e.g. \cite{Charrier_2019, Auger_gal} and briefly summarized below.
    
    The spatial and frequency-dependent galactic brightness temperature $T(\uptheta, \upphi)$ is first obtained from LFMap~\cite{polisensky_2007}. At each point on the sky, the spectral power density per unit solid angle $\Omega$ and per frequency $\nu$, $P_{\nu,\Omega}$ can be computed as:
    \begin{align}
    P_{\nu,\, \Omega}(\uptheta, \upphi) &= \tfrac{1}{2} \, B_{\nu, \, \Omega}(\uptheta, \upphi) \, A_{\rm eff}(\uptheta, \upphi), \label{eq:power}\\
    B_{\nu,\, \Omega}(\uptheta, \upphi) &= \frac{2 \nu^2 k_B T(\uptheta, \upphi)}{c^2}.
    \end{align}
    where we define the effective area of antenna arm $i$ as $A_{\rm eff}(\uptheta, \upphi) = \Vert \rbm{l}_{\mathrm{eff},\, i} \Vert^2$. $B_{\nu, \Omega}$ is the spectral radiance, representing the received power per unit frequency, area, and solid angle. The factor $\tfrac{1}{2}$ in equation~\ref{eq:power} results from the fact that the emission is not polarized, thus only half of the power aligns with $\rbm{l}_{\rm{eff}}$. An illustration of the effective area of the X arm and the spectral radiance $B_{\nu, \Omega}$ is shown in Figure~\ref{fig:galatic_emission}.
    
    Integrating this power over the full sky yields the total spectral power $P_\nu$ received at each frequency. This power corresponds to an open-circuit voltage spectral density on the antenna given by:
    \begin{equation}
    V^2_{\rm{OC,\,RMS}}(\nu) = \frac{P_\nu}{Z_0},
    \end{equation}
    where $Z_0$ is the impedance of free space. The contribution is then filtered through the RF chain response (as previously modeled) to obtain the output root-mean-square voltage per frequency unit:
    \begin{equation}
    V^2_{\rm{RMS}}(\nu) = |\mathbfcal{R}(\nu)|^2 \cdot V^2_{\rm{OC,\,RMS}}(\nu).
    \end{equation}
    
    To generate time-domain noise traces, we finally construct the Fourier transform of a random signal with real and imaginary part sampled from a centered Gaussian with variance:
    \begin{equation}
        V[\Re(\Tilde{\rm V}_{\rm out})]=V[\Im(\Tilde{\rm V}_{\rm out})] =\frac{N f_s}{4} \cdot V^2_{\rm RMS},
    \end{equation}
    where $N$ is the number of time samples and $f_s$ the sampling frequency.
    
    In remote areas such as the GP300 site, Galactic emission is the dominant source of environmental noise\,\cite{GRANDHW}, except for specific communication frequencies which can be efficiently removed by digital notch filters. Besides, the electronics --and preamplifier in particular-- are designed to be a minor contribution to the the stationary noise\,\cite{XuXing}. Finally, radio-frequency interference (RFI) transient signals are too scarce (typical rate below 1\,kHz for GRANDProto300) to contribute to the total power spectrum or be in coincidence with air-shower pulses. All in all, the computed Galactic noise contribution provides an excellent first-order approximation of the noise conditions for radio experiments such as GRANDProto300.
    
    \item Second, we account for the finite timing resolution of the detector by adding a Gaussian uncertainty with $\sigma = 5$\,ns to the trigger times of all simulated signals, following~\cite{Decoene_2023}. Trigger times $\rbm T$ are extracted from the quadratic sum of the X and Y voltage time traces and are defined as the position of the maximum of the corresponding Hilbert envelope. The signal amplitude is defined as the maximum value of this envelope. We disregard the upward component of the voltage signal, because this channel is noisier due to RFI and EAS emissions are predominantly in the horizontal components at the simulation location. 
    
    \item Finally, we account for systematic antenna calibration uncertainties by introducing to the total signal a multiplicative smearing factor sampled from a Gaussian distribution centered at 1 with a standard deviation of 7.5\%, a value consistent with the systematic error for the AERA experiment\,\cite{AERA_calib}.  
\end{itemize}

To emulate realistic triggering conditions, antennas whose peak amplitude is below $5\sigma_{\rm N}$ are discarded, where $\sigma_{\rm N} = \sqrt{2}\,\sigma_{\rm noise}$ and $\sigma_{\rm noise}$ is the standard deviation of the noise per channel, typically around 1.0\,mV in our treatment. The $\sqrt{2}$ factors arise from taking the trigger on the two transverse components. 
This cut ensures that only realistic detections are retained and that the extracted peak values are not dominated by noise fluctuations. Additionally, only events with at least five antennas passing the amplitude cut are kept, which is a minimal requirement for reliable reconstruction.

After the full selection procedure is applied, 7,975 simulated events are retained from the initial set of 25,000. The median number of triggered antennas per event after the cuts is 20.

\section{Graph neural networks on distributed radio antenna arrays}
\label{section:methods}
\subsection{Motivations for using graph neural networks}

Radio arrays designed to detect ultra-high-energy cosmic rays (UHECRs) are typically composed of large numbers of antennas distributed over wide areas. When an extensive air shower (EAS) occurs, it triggers only a subset of these antennas, whose number and spatial distribution can vary significantly depending on the shower energy, direction, and core position.

This variability presents several challenges for both standard reconstruction methods and traditional machine-learning approaches. Convolutional Neural Networks (CNNs), for example, generally rely on data arranged on regular orthogonal grids, which requires dedicated preprocessing or training strategies to accommodate detector geometries with irregular or incomplete sampling~\cite{IceCube_CNN}. Moreover, adapting CNNs to variable trigger patterns often requires mapping the full detector geometry onto fixed-size representations, introducing unnecessary computational overhead.


Graph Neural Networks (GNNs), on the contrary, naturally handle irregular detector geometries. They operate on graph-structured data, where nodes represent entities (here, antennas) and edges encode relationships between them, such as spatial proximity.

\subsection{Wavefront analytical reconstruction}
The machine-learning approach considered in this paper uses a first estimate of the direction to improve the reconstruction. For this purpose, we assume a planar wavefront (PWF) for the radio signal and use the semi-analytical solver described in~\cite{Ferriere2024} to compute the reconstructed plane wave direction $\rbm{k}_{\rm PWF}$ and associated arrival times $\rbm{T}_{\rm PWF}$. 


\subsection{Preprocessing for GNN input}

To construct graph-structured data suitable for GNN processing, we follow a systematic preprocessing pipeline. For each simulated event, a graph is generated by connecting every triggered antenna (node) to its eight nearest neighbors. This configuration was found to provide the best model performance: using fewer neighbors reduces graph connectivity, while increasing the number of neighbors does not lead to significant performance improvements and increases the computational cost.

\begin{figure}
    \centering
    \includegraphics[width=0.5\textwidth]{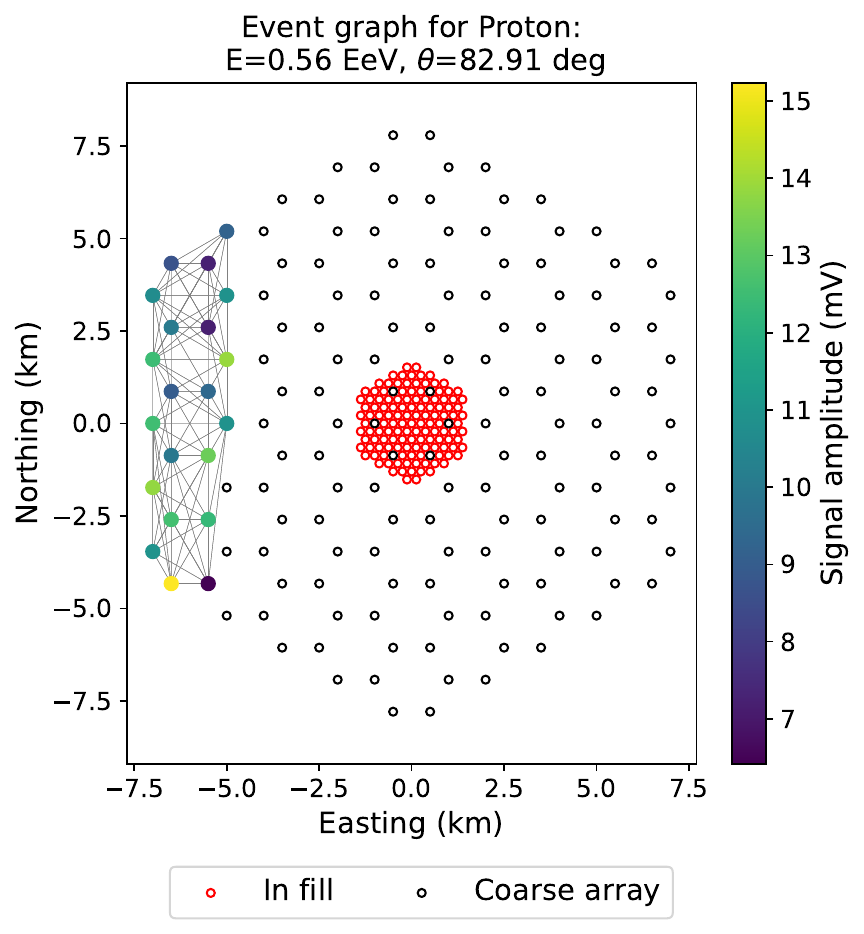}
    \caption{Example of a graph constructed from a simulated event. Triggered antennas are represented as nodes (blue dots), with edges (gray lines) connecting each antenna to its eight nearest neighbors. The color of each node corresponds to the signal amplitude recorded at that antenna, illustrating the spatial distribution of the detected signals across the array.}\label{fig:graph_exemple}
\end{figure}

With eight neighbors per node and the number of graph layers used in the model (see next section), information can propagate across the full graph. This may explain why increasing the connectivity beyond this value does not improve the reconstruction performance. An example of an event mapped as a graph is shown in Fig.~\ref{fig:graph_exemple}.

Each node is associated with a set of input features. These features are explicitly derived from the simulated event and detector geometry:

\begin{itemize}

    \item The \textbf{3D position} of the antenna, $(X, Y, Z)$, known precisely from the detector layout. The X axis points toward the geomagnetic north, the Y axis toward the east, and the Z axis upward. The center of the array is taken as the origin of the coordinate system.

    \item The \textbf{time of arrival} of the radio signal at  antenna positions $\rbm T$, extracted from the simulated time traces following the method described in Section~\ref{sec:noise}. This quantity encodes information about the wavefront geometry and is widely used in standard reconstruction methods.

    \item The \textbf{signal amplitude}, $A$, extracted from the peak of the Hilbert envelope of the voltage trace. The spatial distribution of amplitudes at ground level carries information about the shower geometry, as the radio footprint is elongated along the shower direction. 
    
    \item The \textbf{PWF timing residuals},
    \[
    \boldsymbol{\Delta}_{\rm PWF} = \rbm T - \rbm T_{\rm PWF},
    \]
    which quantify the deviation of the wavefront from a purely planar geometry.

\end{itemize}

In addition, the \textbf{PWF incoming direction}, represented by the unit vector $\rbm{k}_{\rm PWF}$ is given as a global feature after the graph convolution has been performed (see next subsection). This allows the model to learn systematic corrections to the PWF reconstruction, which provides a first-order approximation of the wavefront geometry. The motivation for such inputs is discussed in~\ref{app:training_size}.

\subsection{Architecture}
\begin{figure*}
    \includegraphics[width=\textwidth]{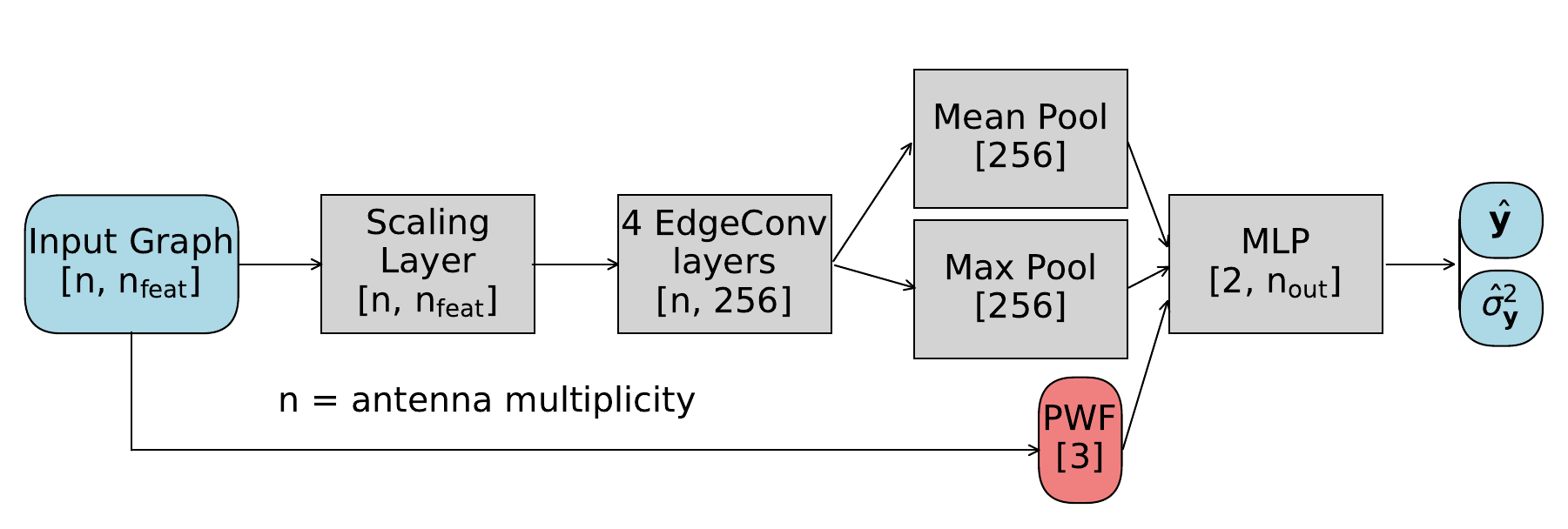}
    \centering
    \caption{Illustration of the architecture used in this study. The model consists of a scaling layer, 4 EdgeConv layer where each node aggregates information from its 8 nearest neighbors into a vector of 256 features, 2 pooling layers that aggregate the node features into a single vector, and a final MLP that outputs the reconstructed quantities and their associated uncertainties. The number of input features $n_{\rm feat}=6$.}\label{fig:GNN_architecture}
\end{figure*}

The general architecture of the GNN is given in Figure \ref{fig:GNN_architecture}.
The model consists of several key components, including a scaling layer, 4 graph convolution layers, 2 pooling layers, and a final MLP that outputs the reconstructed quantities and their associated uncertainties.

The scaling layer forces derived inputs to be distributed with comparable statistics, typically close to a centered and normalized Gaussian. Among the input features, the amplitude distribution follows a power law between 6~mV (corresponding to the 5$\sigma_{\rm noise}$ threshold) and 300~mV (maximal amplitudes around 20$\sigma_{\rm noise}$) while the distribution of timings is uniform with values ranging from 0 to 100~$\mu$s. This helps the model to converge faster and avoid some numerical issues. 

The EdgeConv~\cite{Edge_conv} layers then computes, for every connected pair of nodes, a non-linear parametrizable function of their features as follows:
\begin{equation}
    \rbm{x}^{(k+1)}_{i} = \frac{1}{|\mathcal{N}_i|}\sum_{j \in \mathcal{N}_i} f^{k+1}_p \!\left(\rbm{x}^{(k)}_{i},\,\rbm{x}^{(k)}_{j}\right),
\end{equation}
$\mathcal{N}_i$ being the set of neighbors of node $i$, $\rbm{x}^{(k)}_{i}$ the node features after layer $k$, and $f^{(k+1)}_p$ a learnable function with parameters $p$. In this work, $f_p$ is implemented as a two-layer multilayer perceptron (MLP), i.e. a fully connected feedforward neural network. Four EdgeConv layers are used in total.

To obtain a fixed-size graph representation, both max-pooling and mean-pooling operations are applied over the node embeddings. For each feature, the maximum and mean values across all nodes are computed and concatenated into a single vector, which is then used as input to a final MLP predicting the target quantities.

For each reconstructed quantity, the model predicts both a mean value and an associated variance. The mean corresponds to the reconstructed parameter, while the variance provides an estimate of the prediction uncertainty under the assumption of Gaussian-distributed residuals. This probabilistic interpretation is enforced during training, as described in the next section.

Finally, for each reconstruction task, we train an ensemble of $N$ models with different random weight initializations in order to improve both prediction accuracy and uncertainty estimation, a critical information to assess the reliability of the reconstructed quantities.

\subsection{Training and inference procedure}\label{section:training_methods}

We train an ensemble of 12 models, which was found sufficient to improve robustness and reduce prediction variance, as no further gains in precision or uncertainty calibration are observed for larger ensemble sizes. The models differ only in their initial weights and training data ordering. To train the model, we need to define a loss function. As we want to predict both a point estimate and an uncertainty, we use a Gaussian negative log-likelihood~\cite{NLL_1994} (NLL) loss function, which encourages well-calibrated predictive distributions and penalizes both biased means and incorrect variance estimates. It is defined for a scalar target $y$ as:
\begin{equation}\label{eq:NLL}
\mathcal{L}\left(\hat{y},\, \hat{\sigma}^2\right)
=
\frac{1}{2}
\left(
\frac{(y-\hat{y})^2}{\hat{\sigma}^2+\epsilon}
+
\log(\hat{\sigma}^2+\epsilon)
\right),
\end{equation}
where $\hat{y}$ is the predicted mean and $\hat{\sigma}^2$ the predicted variance associated with the target $y$. A small constant $\epsilon = 10^{-6}$ is added to improve numerical stability.

This loss function can be viewed as a probabilistic generalization of the classical mean squared error (MSE) loss. It encourages the model to produce accurate predictions while simultaneously learning meaningful uncertainty estimates.

Each model is trained using the Adam optimizer with adaptive learning rates to ensure efficient convergence. A batch size of 4 is used for training. The dataset is split into 5000 events for training and 1200 events for validation. All models are trained on the same set of events. This split remains identical for all models to ensure comparability; only the initial weights and the input data order are modified. Training is conducted for a maximum of 120 epoch which always leads to convergence of the models. Early stopping was also considered to stop the training if no improvement is observed in the validation loss over a predefined number of epochs.

Each model $f_i$ (where $i = 1, 2, \dots, 12$) is trained independently for predicting the best estimate $\hat{y}_i$ and its associated uncertainty $\hat{\sigma}_{y,\,i}$ for a given quantity. The ensemble prediction $\hat{y}$ is computed as the mean of the individual model predictions:
\[
\hat{y} = \frac{1}{N} \sum_{i=1}^{N} \hat{y}_i,
\]
where $N = 12$ is the number of models in the ensemble.

The associated uncertainty $\hat{\sigma_{y}}$ is estimated by combining the predicted uncertainties $\hat{\sigma}_{y,\,i}$ from each model and the variance of the predictions:
\begin{equation}\label{eq:ensemble_uncertainty}
    \hat{\sigma_y}^2 = \frac{1}{N} \sum_{i=1}^{N} \hat{\sigma}_{y,\,i}^2 + \frac{1}{N} \sum_{i=1}^{N} {\left( \hat{y_i} - \hat{y}\right)}^2.
\end{equation}
This formulation captures both aleatoric uncertainty through the individual model uncertainties combined (left term) and epistemic uncertainty (model uncertainty due to limited data or model capacity) through the variance of the ensemble predictions (right term).

\subsubsection*{Direction reconstruction}
For direction reconstruction, we predict the three Cartesian components of the unit direction vector
\[
\rbm{k} = (k_x, k_y, k_z),
\]
defined in terms of the zenith angle $\theta$ and azimuth angle $\phi$ as

\begin{equation}\label{eq:direction_vector}
\rbm{k} = -
\begin{pmatrix}
\sin\theta \cos\phi \\
\sin\theta \sin\phi \\
\cos\theta
\end{pmatrix}.
\end{equation}

This representation is convenient for GNN training because it avoids the coordinate singularities associated with angular parametrizations and allows for straightforward computation of angular distances. It also naturally enables the use of loss functions based on Euclidean distances. The predicted zenith and azimuth angles can subsequently be recovered by inverting Eq.~\ref{eq:direction_vector}.

For direction reconstruction, the loss function is given by Eq.~\ref{eq:NLL}, applied independently to each Cartesian component of the direction vector.

This formulation does not explicitly enforce the unit-norm constraint on $\rbm{k}$. Instead, the predicted vector is normalized at inference time. In practice, this does not significantly affect the training procedure, as the model naturally learns predictions close to the unit sphere.

Alternative loss functions were also investigated. In particular, the mean squared error (MSE) loss yielded similar reconstruction performance but did not provide calibrated uncertainty estimates, even when combined with ensemble methods.

\subsubsection*{Energy reconstruction}
For energy reconstruction, the Gaussian-NLL loss function applied to the log energy was chosen as it naturally handles the wide dynamic range of energies. The loss function for energy reconstructioncan also be given by equation~\ref{eq:NLL} where $y$ stands for $\log E$, E being the true electromagnetic energy of the shower. \newline

\section{Performances}\label{section:performances}
Before evaluating the performance of the models, we first assess their convergence during training. We then present the results for direction and energy reconstruction, focusing first on accuracy. We assess the uncertainty estimation and calibration in the next section. In the following, we define as \emph{single model} the individual models that make the \emph{ensemble model}. We will discuss the performances of both to demonstrate the necessity of the ensemble. The performance of the single models is the average performance of the individual models.

\subsection{Model convergence}
To evaluate model convergence, we focus on the evolution of the mean squared error on both the training and validation sets over training epochs.

\begin{figure}[t]
    \centering
    \includegraphics[width=0.65\textwidth]{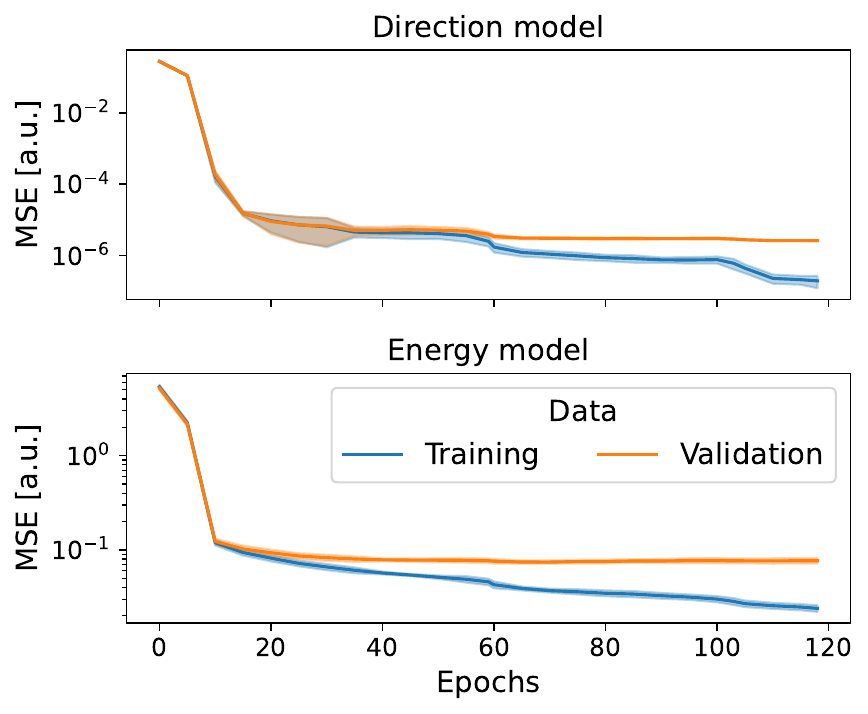}
    \caption{Evolution of the mean squared error on the training and validation sets over epochs for the 12 ensemble members GNN for direction (top) and energy (bottom) reconstruction. The validation error stabilizes  early in both cases. The transparent envelopes represent the standard deviation over all 12 models}
    \label{fig:loss_evolution}
\end{figure}

To assess training stability and potential overfitting, we monitor the error across the 12 ensemble members (see Figure~\ref{fig:loss_evolution}). The mean squared error is shown for both the training set (blue) and validation set (orange), while the transparent envelopes indicate the standard deviation over all ensemble members. All runs converge smoothly, with no significant improvement observed after approximately 20 epochs, reaching consistent performance on both training and validation sets.

 We do see that the validation errors stabilize early on in the training 
 while the performance on training set keeps improving, a possible hint for mild overfitting. In all training curves, step-like decreases followed by plateaus are observed. These correspond to successive reductions of the learning rate triggered by the learning-rate scheduler when no improvement is observed on the validation loss.

We further observe that incorporating PWF information significantly reduces both training and validation losses for direction reconstruction. In contrast, for energy reconstruction, the inclusion of PWF inputs has little impact on the evolution of the MSE.

We also investigate data efficiency by training on sub-sampled datasets (25\%, 50\%, and 75\% of the full training set). All configurations converge, with degraded performance at smaller training sizes. For direction reconstruction, physics-informed models (including PWF residuals) are more sample-efficient, achieving better performance with fewer events, while purely simulation-driven models benefit more from larger datasets but do not reach the same level of precision. This highlights the benefit of incorporating domain knowledge to guide learning, particularly in low-data regimes (see~\ref{app:training_size}). This effect is less pronounced for energy reconstruction, for which the provided physical inputs are not directly correlated with the target variable.

\subsection{Performances on direction reconstruction}
Constructing the ensemble model as described in Section~\ref{section:training_methods} yields a robust direction estimator with associated uncertainties. The angular resolution is commonly defined as the median 
angular error between the true and reconstructed directions, where the angular error is defined as
\begin{equation}
\Psi = \arccos\left(\rbm{k} \cdot \hat{\rbm{k}}\right)\, .
\end{equation}

\begin{figure}[t]
    \centering
    \includegraphics[width=0.6\textwidth]{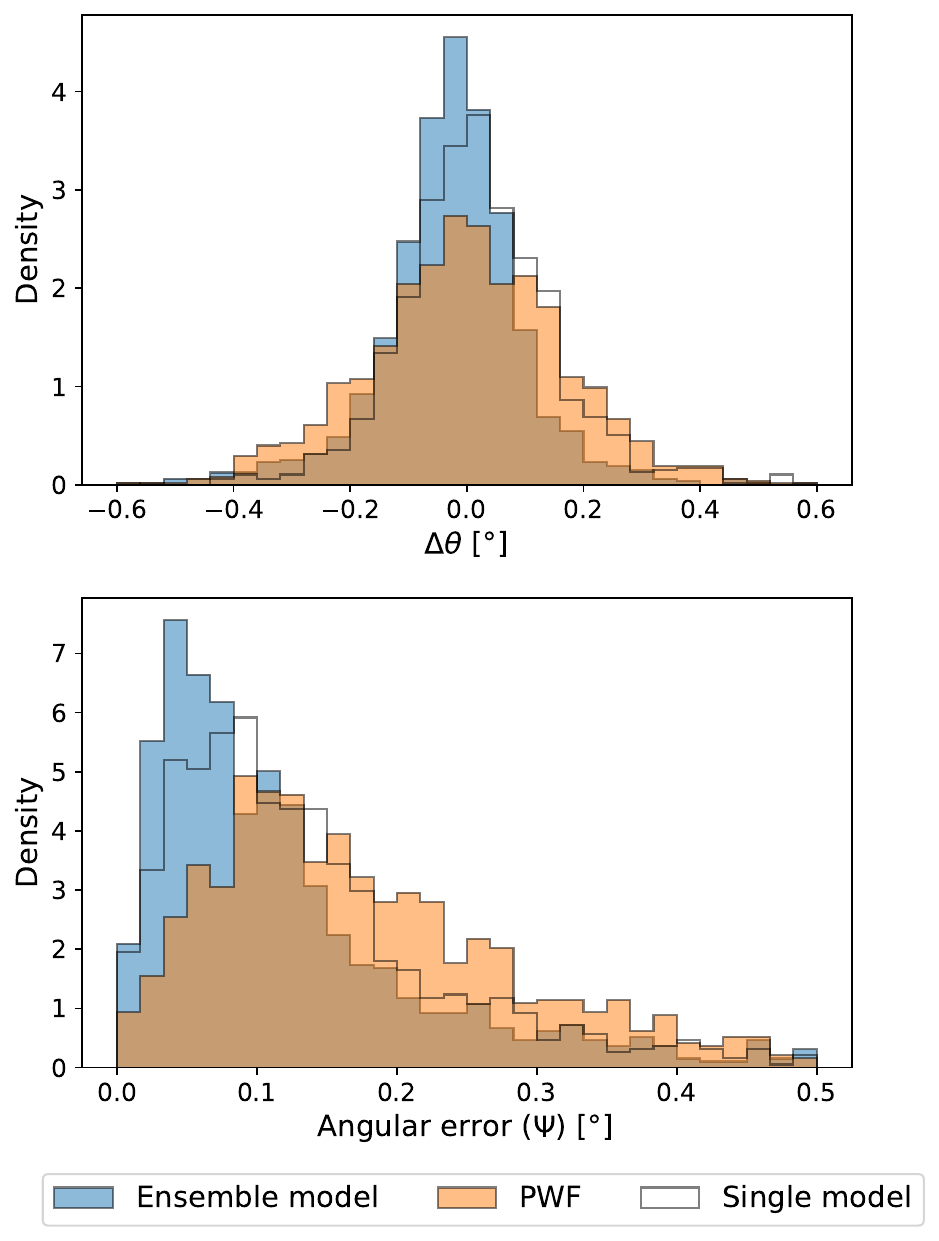}
    \small
    \begin{tabular}{lcccc}
\hline
Method & $\langle\Delta\theta\rangle$ [°] & $\sigma_{\Delta\theta}$ [°] & $q_{50,\Uppsi}$ [°] & $q_{95,\Uppsi}$ [°] \\
\hline
PWF & 5.8×10$^{ -4 }$ & 1.9×10$^{ -1 }$ & 1.6×10$^{ -1 }$ & 4.5×10$^{ -1 }$ \\
Single & 2.1×10$^{ -2 }$ & 1.5×10$^{ -1 }$ & 1.1×10$^{ -1 }$ & 4.1×10$^{ -1 }$ \\
Ensemble & -2.8×10$^{ -2 }$ & 1.4×10$^{ -1 }$ & 9.2×10$^{ -2 }$ & 3.8×10$^{ -1 }$ \\
\hline
\end{tabular}
    \caption{Distribution of angular errors for single models (gray), the ensemble mean prediction (blue), and the PWF estimator (orange). The ensemble model achieves a mean angular error of 0.092°, significantly outperforming both single models (0.11°) and the PWF baseline (0.16°). The table summarizes the performances of all models considered. The quantities shown are the mean error on zenith angle, the standard deviation of the error on zenith angle, the mean angular error, and the 95th percentile of the angular error distribution.
    }\label{fig:angular_resolution}
\end{figure}

We show in Figure~\ref{fig:angular_resolution} the distribution of angular errors for single models, the ensemble mean prediction, and the PWF baseline. The ensemble mean achieves a median angular error of $0.092^\circ$, outperforming both single models ($0.11^\circ$) and the PWF baseline ($0.16^\circ$), demonstrating the gain from both ensembling and physics-informed inputs.

For inclined events, reconstruction is generally more accurate for azimuth than for zenith because of the better lever arm in the lateral direction A small bias is observed in $\theta$, but its magnitude remains significantly smaller than the overall resolution (see Figure\,\ref{fig:angular_resolution}).

\begin{figure}[t]
    \centering
    \includegraphics[width=.8\textwidth]{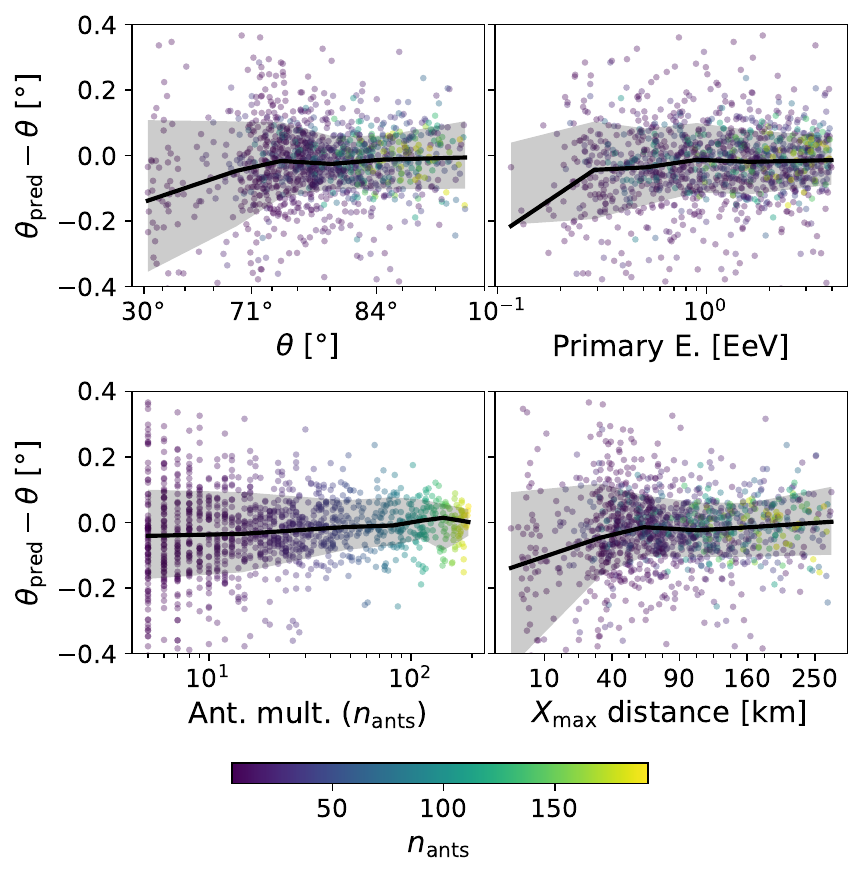}
    \caption{Zenith-angle residual ($\Delta \uptheta$) versus event parameters for the ensemble model\@. Top-left: zenith; top-right: primary energy; bottom-left: triggered-antenna multiplicity; bottom-right: $X_{\max}$ distance to the array center. Performance depends mainly on multiplicity; no significant bias is observed. The shaded band indicates the central 68\% interval.}\label{fig:res_theta_vs_quantities}
\end{figure}

The angular resolution strongly depends on the number of triggered antennas, as fewer stations reduce both timing constraints and geometric lever arm. This is illustrated in Figure~\ref{fig:res_theta_vs_quantities} on the residual in $\theta$, as it is the major contribution to the total angular error. Since the number of triggered antennas correlates with both energy and zenith angle, the resolution also exhibits a mild dependence on these parameters. For very inclined showers, although more antennas are triggered because of the larger footprint size, the $\theta$ resolution doesn't improve because of the modest antenna height variation over the layout, providing a moderate lever arm for zenith angle reconstruction. 

\begin{figure}
    \centering
    \includegraphics[width=0.9\textwidth]{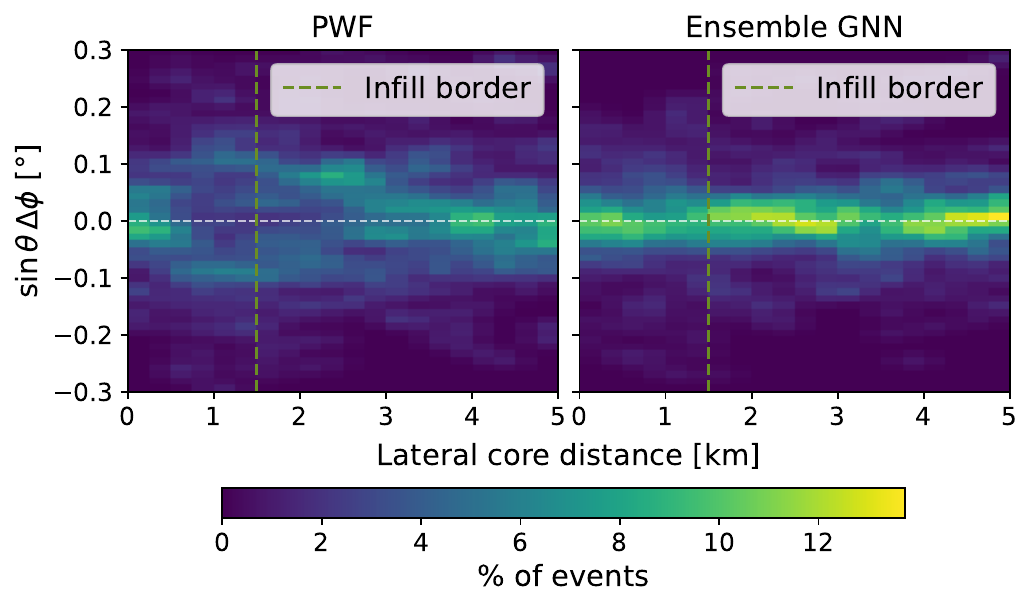}
    \caption{Conditional distribution of $\sin\theta\,\Delta\phi$ versus lateral core distance for the PWF (left) and ensemble GNN (right), with each column normalised to 100\%. The lateral core distance corresponds to the distance between the shower core and the shower axis projected onto the ground. The vertical line correspond to the edge of the infill. The bi-modality introduced by the PWF for events on the edge of the infill is completely corrected by the GNNs} \label{fig:angular_distance_on_phi}
\end{figure}

The shower core distance also affects the PWF angular resolution. We show in Figure~\ref{fig:angular_distance_on_phi} that for distant events, the sampling of the shower wavefront by the detector becomes asymmetric ---more antennas being on one side of the shower than the other--- which introduces a bias in the plane-wave reconstruction. This effect is significantly reduced with the ensemble model, which maintains a stable angular resolution across core distances. 

Overall, the ensemble model achieves robust performance across a wide range of event topologies. No significant systematic bias is observed across the explored parameter space, demonstrating its effectiveness for direction reconstruction in radio antenna arrays.

\subsection{Performances on energy reconstruction}
As the reconstructed quantity for energy is the logarithm of the energy, the ensemble prediction is obtained as the mean of the predicted log-energies. This is equivalent to taking the geometric mean of the individual model predictions:
\begin{equation}
\hat{E}
=
\exp\left(\frac{1}{N}\sum_{i=1}^{N}\log \hat{E}_i\right)
=
\left(\prod_{i=1}^{N}\hat{E}_i\right)^{1/N}.
\end{equation}

The energy reconstruction performance is evaluated using the relative error:
\begin{equation}
\frac{\Delta E}{E} = \frac{E - \hat{E}}{E}.
\end{equation}

The energy resolution $\rm R_{E,\,68}$, commonly defined as the half-width of the 68\% central interval of the relative error distribution, i.e. 
\begin{equation}
{\rm R_{E,\,68}} = \tfrac{1}{2}\!\left(Q_{84\%}\!\left(\tfrac{\Delta E}{E}\right)-Q_{16\%}\!\left(\tfrac{\Delta E}{E}\right)\right)
\end{equation}

\begin{figure}[t]
    \centering
    \includegraphics[width=0.65\textwidth]{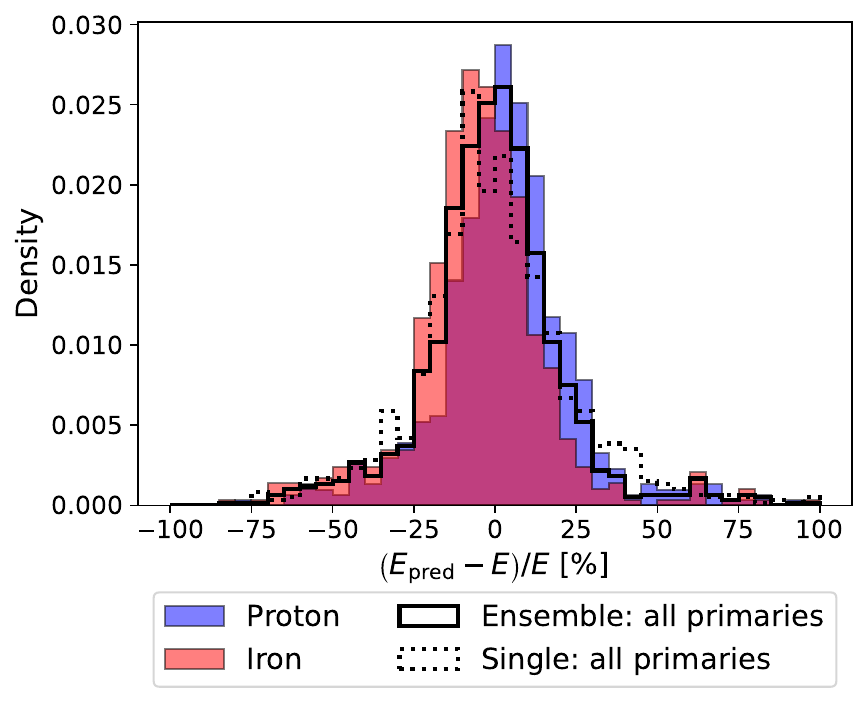}
    \small
    \begin{tabular}{lcccc}
\hline
Method & $\langle  \Delta \rm{E} / \rm{E}\rangle$ [\%] & ${\rm R_{E,\,68}}$ [\%] & ${\rm R_{E,\,95}}$ [\%]\\
\hline
Proton & 4.4 & 16.0 & 51.4 \\
Iron & -4.4 & 15.3 & 51.2 \\
All prim.: Ensemble& 0.1 & 16.4 & 51.3 \\
All prim.: Single model& 0.7 & 19.5 & 54.2 \\
\hline
\end{tabular}
    \caption{Distribution of relative energy errors $(E_{\rm pred}-E)/E$. Blue and orange filled histograms: ensemble predictions for proton and iron primaries, respectively. The ensemble model achieves an energy resolution of 16.4\% (68\% quantile of the absolute relative residuals), compared with 19.5\% for single models. A small bias is visible in the ensemble: iron events are slightly underestimated while proton events are slightly overestimated, but resolution is the same.}\label{fig:energy_resolution}
\end{figure}

Figure~\ref{fig:energy_resolution} shows the distribution of relative errors for single models and for the ensemble prediction. The ensemble achieves an energy resolution of 16.4\%, compared to 19.5\% for single models. The mild improvement is mainly driven by the suppression of outliers, which are more pronounced in individual models.

A small composition-dependent bias is observed: iron primaries tend to be slightly underestimated, while proton primaries are slightly overestimated. This likely reflects differences in shower development and radio emission properties between primary types that are not fully captured by the model. However, the effect remains small compared to the overall resolution.

\begin{figure}[t]
    \centering
    \includegraphics[width=0.8\textwidth]{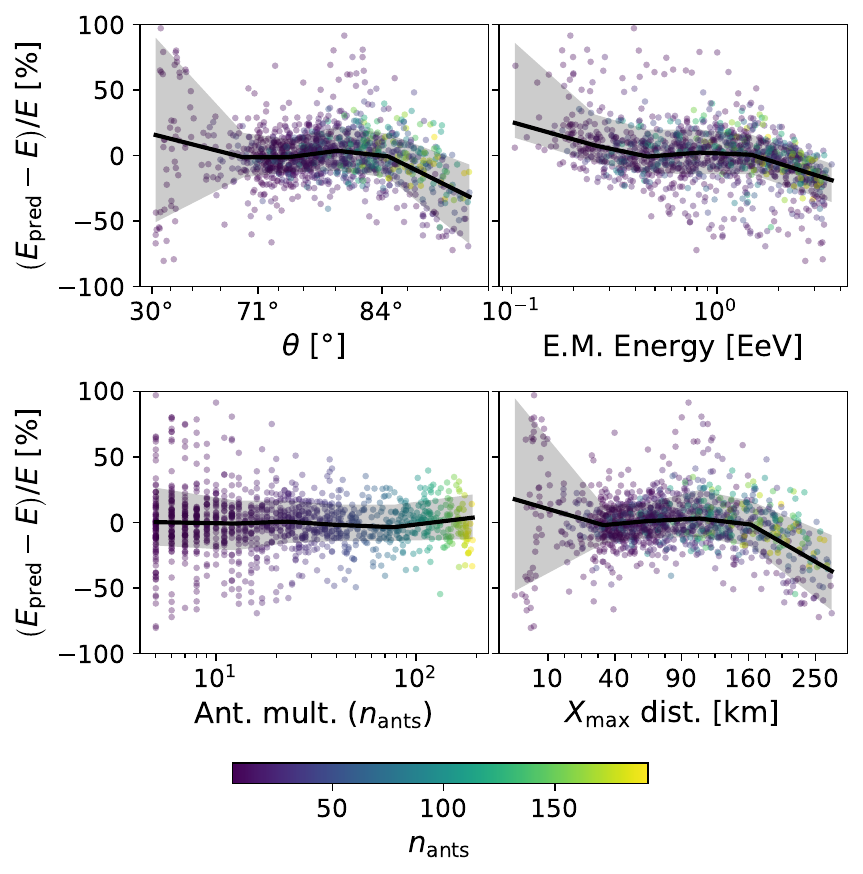}
    \caption{Relative energy error as a function of event parameters for the ensemble model\@. Top-left: zenith angle; top-right: electromagnetic energy; bottom-left: number of triggered antennas; bottom-right: distance from $X_{\max}$ to the array. Blue points denote individual events; the black curve shows the bin-wise mean with a shaded 68\% envelope, representing the energy resolution as a function of the horizontal axis. Bias is observed at high zenith angles corresponding to large distances of the emission point.}\label{fig:energy_resolution_vs_quantities}
\end{figure}

The dependence of the energy resolution on event properties is shown in Figure~\ref{fig:energy_resolution_vs_quantities}. The resolution improves with increasing antenna multiplicity, reflecting the larger amount of timing and amplitude information available for reconstruction. A mild dependence on zenith angle is also observed, with a tendency to underestimate the energy at large zenith angles. Such showers develop farther from the detector, leading to reduced signal coherence and weaker radio amplitudes, which may contribute to an underestimation of the reconstructed energy, as pointed out in ~\cite{chiche_decoherence}.

A similar trend is observed as a function of the distance between $X_{\rm max}$ and the array, which is strongly correlated with zenith angle. 

Importantly, the energy resolution remains stable across the full energy range considered, demonstrating that the model can handle the large dynamic range of cosmic-ray energies. The distance of the shower core to the array center has only a mild impact, with slightly improved performance for events closer to the array where antenna density is higher. Overall, the ensemble model provides robust energy reconstruction across a wide range of event characteristics.

\section{Uncertainty estimation}\label{sec:uncertainty_estimations}
Beyond reconstructing the shower direction and electromagnetic energy, estimating the associated uncertainties is essential for assessing the reliability of reconstructed cosmic-ray events.

To this end, prediction residuals are modeled as Gaussian-distributed quantities, and the network is trained using a Gaussian negative log-likelihood loss function (see Section~\ref{section:training_methods}). Combined with deep ensembling, this approach provides event-by-event uncertainty estimates alongside the reconstructed quantities.

\subsection{Uncertainty propagation}
For direction reconstruction, the uncertainty on the three spatial coordinates of $\rbm{k}$ is propagated to the uncertainty on the angles $\uptheta$ and $\upphi$. Following the derivations from~\ref{app:propa_theta_phi} we obtain:
\begin{equation}
    \left(\begin{matrix}\uptheta \\ \upphi\end{matrix}\right)
    \sim
    \mathcal{N}\left(  \left(\begin{matrix} \hat{\uptheta} \\ \hat{\upphi} \end{matrix}\right),\,
    \hat{\bm{\Sigma}}_{\uptheta,\upphi}
    \right)
\end{equation}
with $\hat{\uptheta}$ and $\hat{\upphi}$ the polar angles of the reconstructed vector $\hat{\rbm{k}}$ and $\hat{\bm{\Sigma}}_{\uptheta,\upphi}$ the covariance matrix of the uncertainty on $\uptheta$ and $\upphi$ given by:
\begin{equation}\label{eq:variance}
    \hat{\bm{\Sigma}}_{\uptheta,\upphi} \;=\;
    \Big{[\, \rbm{R_a}^{\!\rm T}\,\hat{\bm{\Sigma}}^{-1}\,\rbm{R_a}\,\Big]}^{-1},\
\end{equation}
where $\bm{\hat{\Sigma}}$ is the predicted covariance matrix on $\rbm{k}$, and $\rbm{R_a}$ is given by:
\begin{equation}
    \rbm{R_a} 
    = \frac{\partial\, \rbm{k} }{\partial\, (\uptheta, \upphi)} 
    = \begin{pmatrix}
            -\cos\uptheta\cos\upphi & \;\;\sin\uptheta\sin\upphi \\
            -\cos\uptheta\sin\upphi & -\sin\uptheta\cos\upphi \\
            \;\;\sin\uptheta        & \;\;0
    \end{pmatrix}.
\end{equation}

From $\hat{\bm{\Sigma}}_{\uptheta,\upphi}$ we extract
\begin{equation}
    \hat{\sigma}_{\uptheta} = \sqrt{{(\hat{\bm{\Sigma}}_{\uptheta,\upphi})}_{11}},\quad
    \hat{\sigma}_{\upphi}   = \sqrt{{(\hat{\bm{\Sigma}}_{\uptheta,\upphi})}_{22}},\quad
    \hat{\rho}_{\uptheta,\upphi} = \frac{{(\hat{\bm{\Sigma}}_{\uptheta,\upphi})}_{12}}{\hat{\sigma}_{\uptheta}\hat{\sigma}_{\upphi}}.
\end{equation}
These equations respectively yield the predicted standard deviation of the uncertainty on $\uptheta$, $\upphi$ and on the correlation between the two quantities.
The propagation from above rely on first-order (local) linearization, which is accurate given our small errors (typically < 0.1°); higher-order effects affect only the tails of the distribution, outside of measurable regions. \\

For energy reconstruction, the model predicts the mean and variance of the logarithmic energy, denoted respectively by
\[
\mu = \log \hat{E}
\qquad \text{and} \qquad
\sigma^2_{\log E}.
\]

With a normal distribution for $\log E$ (see next section), 
\[
\log E \sim \mathcal{N}(\mu,\sigma^2_{\log E}),
\]
the corresponding moments in linear energy space follow a log-normal distribution:
\begin{align}
    \mathbb{E}[E] &=
    \exp\!\left(\mu + \tfrac{1}{2}\sigma^2_{\log E}\right),\\
    \mathrm{Var}(E) &=
    \left(\exp(\sigma^2_{\log E}) - 1\right)
    \exp\!\left(2\mu + \sigma^2_{\log E}\right).
\end{align}

Confidence intervals are computed in log-space and subsequently exponentiated, which preserves the symmetry and calibration properties of the predictive distribution.

\subsection{Uncertainty calibration}
The quality of the uncertainty calibration is evaluated using two standard diagnostics.

First, we inspect the standardized residuals for all reconstructed quantities. For a target quantity $y$ with predictive mean $\hat{y}$ and predictive standard deviation $\hat{\sigma}_y$, the standardized residual is defined as
\begin{equation}
z = \frac{y - \hat{y}}{\hat{\sigma}_y}.
\end{equation}

\begin{figure}[t]
    \centering
    \includegraphics[width=0.66\textwidth]{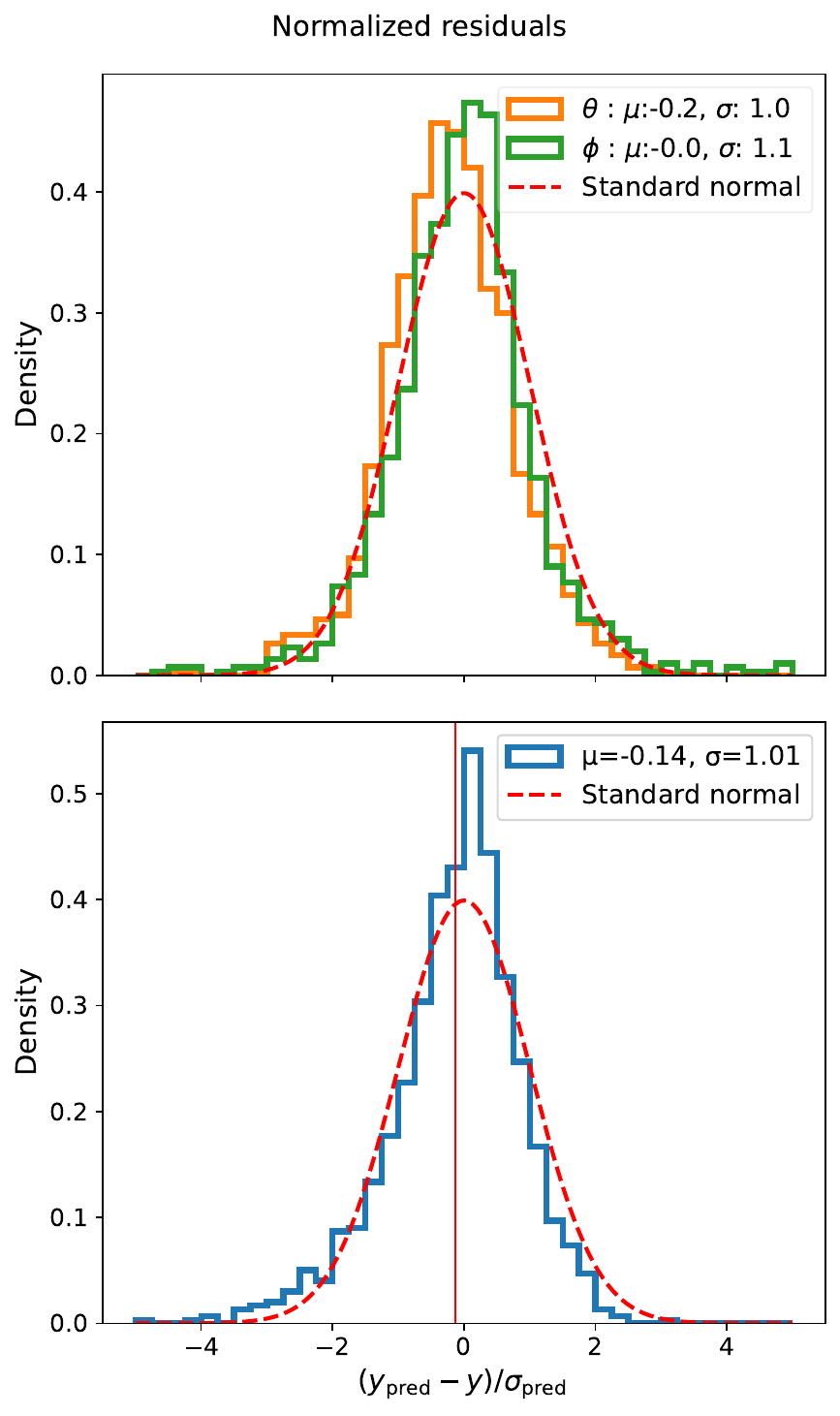}
    \caption{Standardized residuals for the two direction components (top: $\uptheta$, $\upphi$) and for $\log E$ (bottom). The red curve is a standard Gaussian distribution $\mathcal{N}(0,1)$. The close match supports the Gaussian assumption.}\label{fig:standardized_residuals}
\end{figure}

For a well-calibrated model, these residuals are expected to follow a standard normal distribution. Figure~\ref{fig:standardized_residuals} shows the empirical distributions of the standardized residuals for $\theta$, $\phi$, and $\log E$ on the validation dataset. Overall, the distributions closely follow the expected standard normal behavior, with only mild deviations in the tails.

Second, we evaluate calibration using coverage plots. For each event, we compute the nominal coverage level
\[
p = \mathrm{erf}\!\left(\frac{|z|}{\sqrt{2}}\right),
\]
which corresponds to the probability for a standard normal variable to lie within $\pm |z|$ of the mean.

The empirical cumulative distribution function (CDF) of these coverage values is then compared to the ideal diagonal relation. A perfectly calibrated model should follow the diagonal, meaning that an $X\%$ confidence interval effectively contains $X\%$ of the events.

\begin{figure}
    \centering
    \includegraphics[width=0.6\textwidth]{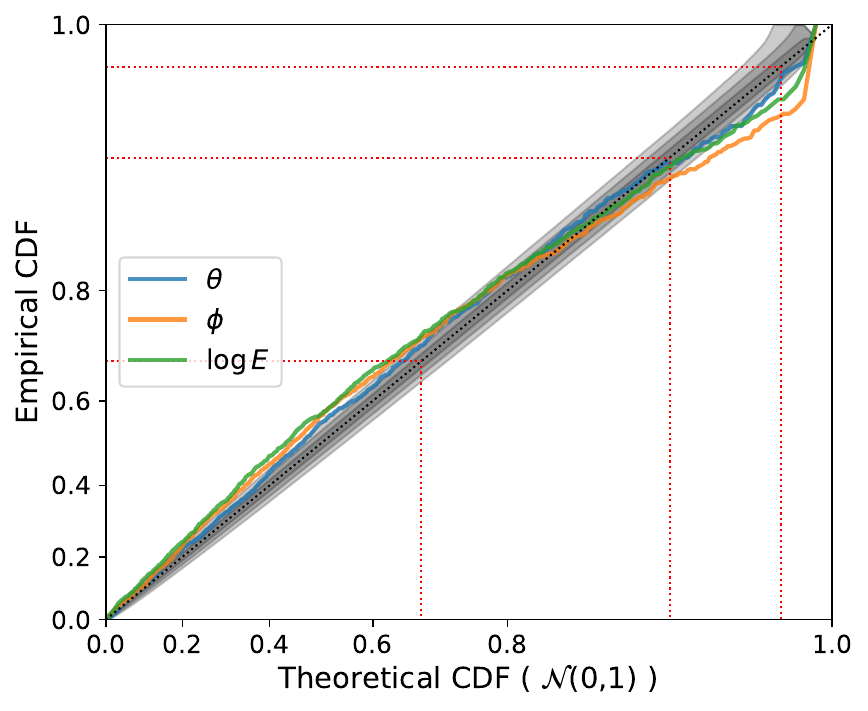}
    \caption{Coverage plots for the zenith angle (orange), azimuth angle (green), and logarithmic electromagnetic energy (blue). The red dotted lines indicate the 68\%, 95\%, and 99.7\% confidence levels expected for a standard normal distribution. The black dashed diagonal corresponds to a perfectly calibrated model, while the gray shaded region represents the expected statistical fluctuations. Overall, the reconstructed quantities follow the expected calibration behavior with only mild deviations.}
    \label{fig:coverage}
\end{figure}

Figure~\ref{fig:coverage} presents the coverage plots for $\theta$, $\phi$, and $\log E$. Curves above the diagonal indicate underconfident predictions (overestimated uncertainties), while curves below the diagonal indicate overconfident predictions (underestimated uncertainties). The black envelope represents the expected statistical fluctuations at the $3\sigma$ level for a perfectly calibrated Gaussian model.

All three reconstructed quantities remain close to the ideal diagonal relation, indicating overall good calibration. For the energy and zenith angle $\theta$, the calibration remains accurate over most confidence levels, with only mild overconfidence in the extreme tails ($\gtrsim 99.7\%$). The azimuth angle $\phi$ exhibits slightly stronger deviations in the tails, although the calibration remains good at lower confidence levels. Overall, the coverage plots indicate that the uncertainty estimates are well calibrated for all reconstructed quantities.

\begin{figure}
    \centering
    \includegraphics[width=.5\textwidth]{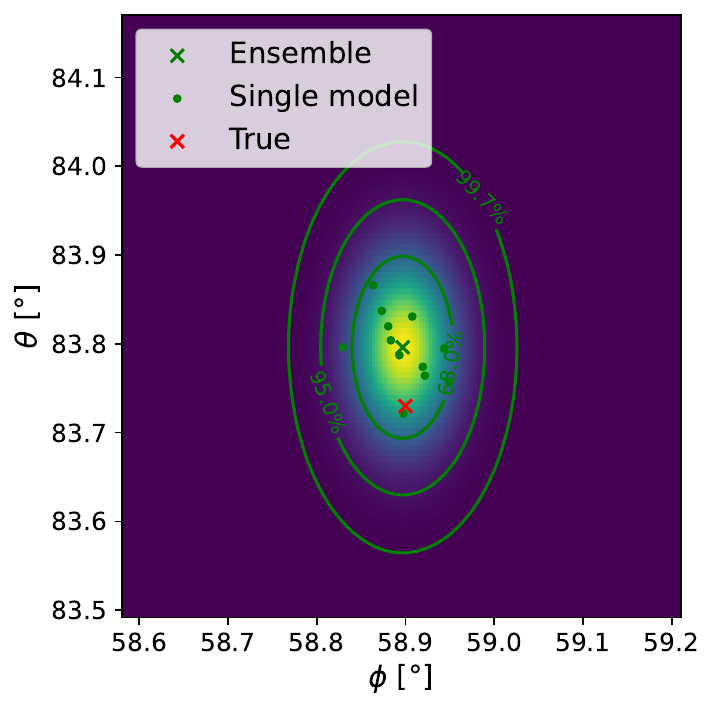}
    \caption{Direction reconstruction with uncertainty estimation. The individual models (green dots) predict a correction to the PWF direction (blue cross), there mean defines the deep ensemble prediction (green cross). The predicted covariance matrix gives the predicted uncertainty distribution which allows to create confidence intervals around the predicted value. Here the true value is inside the "1$\sigma$" interval (68\% isoline). }\label{fig:reconstructed_event}
\end{figure}

Now that resolution and uncertainty calibration are quantified, we can perform reconstruction of events with reliable confidence regions around the prediction. For direction reconstruction, this can be illustrated in Fig.~\ref{fig:reconstructed_event}, where the true shower direction is indicated by a red cross. The individual ensemble members are represented by green markers, while the ensemble prediction is shown as a green cross. Compared to the Plane Wave Fit (PWF) reconstruction shown in blue, the ensemble prediction is both more accurate and associated with a well-defined confidence region derived from the ensemble uncertainty estimates.

\section{Robustness to setup variations}\label{sec:robustness}
In this section, we evaluate the robustness of the trained models to variations in the experimental setup. This is crucial for real-world applications, where conditions may differ from those in the training simulations. We consider three main types of variations: increased threshold levels, variation in antenna gain arising from miscalibration, changes in number of active antennas.

\subsection{Increased threshold levels}
In the standard treatment, the networks are trained on a set of events where the signal amplitude on every antenna is above a 5$\sigma_{\rm N}$ threshold, as it is the trigger condition (see Section~\ref{sec:noise}). In practice, the trigger logic is more complex, particularly in the context of autonomous radio detection~\cite{GRANDProtoPapers_2025}, and actual trigger threshold may therefore be larger than this value. In the following, we therefore assess the impact of a more conservative triggering threshold on the performance of the GNN.

We evaluate the ensemble model performance on datasets constructed with 5 different amplitude thresholds between 5$\sigma_{\rm N}$ and 9$\sigma_{\rm N}$, resulting in a number of reconstructed events between 1200 events at 5$\sigma_{\rm N}$ to 867 events at 9$\sigma_{\rm N}$.

\begin{figure}
    \centering
    \includegraphics[width=0.65\textwidth]{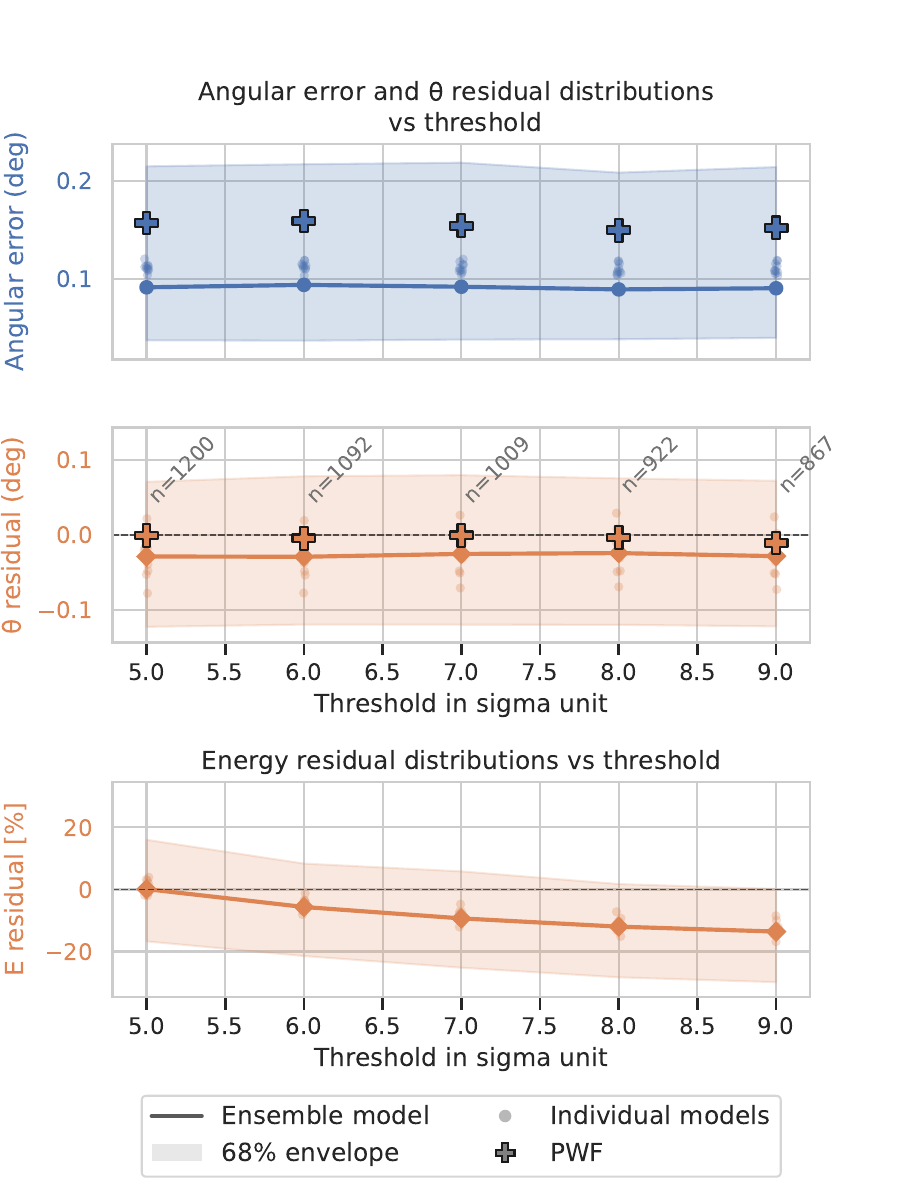}
    \caption{Top: Angular resolution (median values) and zenith angle bias (mean values) as a function of trigger threshold for the ensemble model (solid line), the PWF estimator (crosses) and single models (transparent dots). The angular resolution and error on $\uptheta$ stay stable. Bottom: Energy resolution as a function of the trigger threshold values for ensemble (solid line) and single models (transparent points). TThe energy resolution stays stable but biases with increasing threshold from 0.1\% at $5\sigma$ to -16\% at 9$\sigma$.}\label{fig:robustness_threshold}
\end{figure}

The results are shown in Fig.~\ref{fig:robustness_threshold}. We observe no significant degradation in direction reconstruction performance as the threshold increases. The robustness of the results can be understood from the fact that increasing the threshold effectively reduces the number of triggered antennas, leading to sparser sampling of the same underlying radio footprint, equivalent at first order to lowering the signal amplitude.

This interpretation is further supported by the energy reconstruction, which shows a systematic underestimation of the reconstructed energy with increasing threshold, while preserving a stable resolution across all configurations.

\subsection{Variation in the number of active antennas}
In real-world scenarios, some antennas may be temporarily offline or experience intermittent malfunctioning. It is therefore important to assess the robustness of the reconstruction performance with respect to antenna dropouts.

To quantify this effect, we construct modified validation datasets by randomly deactivating 5\%, 10\%, 20\%, 30\%, and 40\% of the antennas in the array geometry. The corresponding antennas are then removed from all events, which reduces the effective detector sampling. This procedure also leads to a reduced number of reconstructible events, since some configurations no longer satisfy the minimum trigger requirement (fewer than five active antennas). The number of available events decreases from 1200 in the full configuration to 969 when 40\% of antennas are deactivated. The performance of the ensemble model, trained on the full detector layout, is then evaluated on these modified datasets.\\

\begin{figure}
    \centering
    \includegraphics[width=0.65\textwidth]{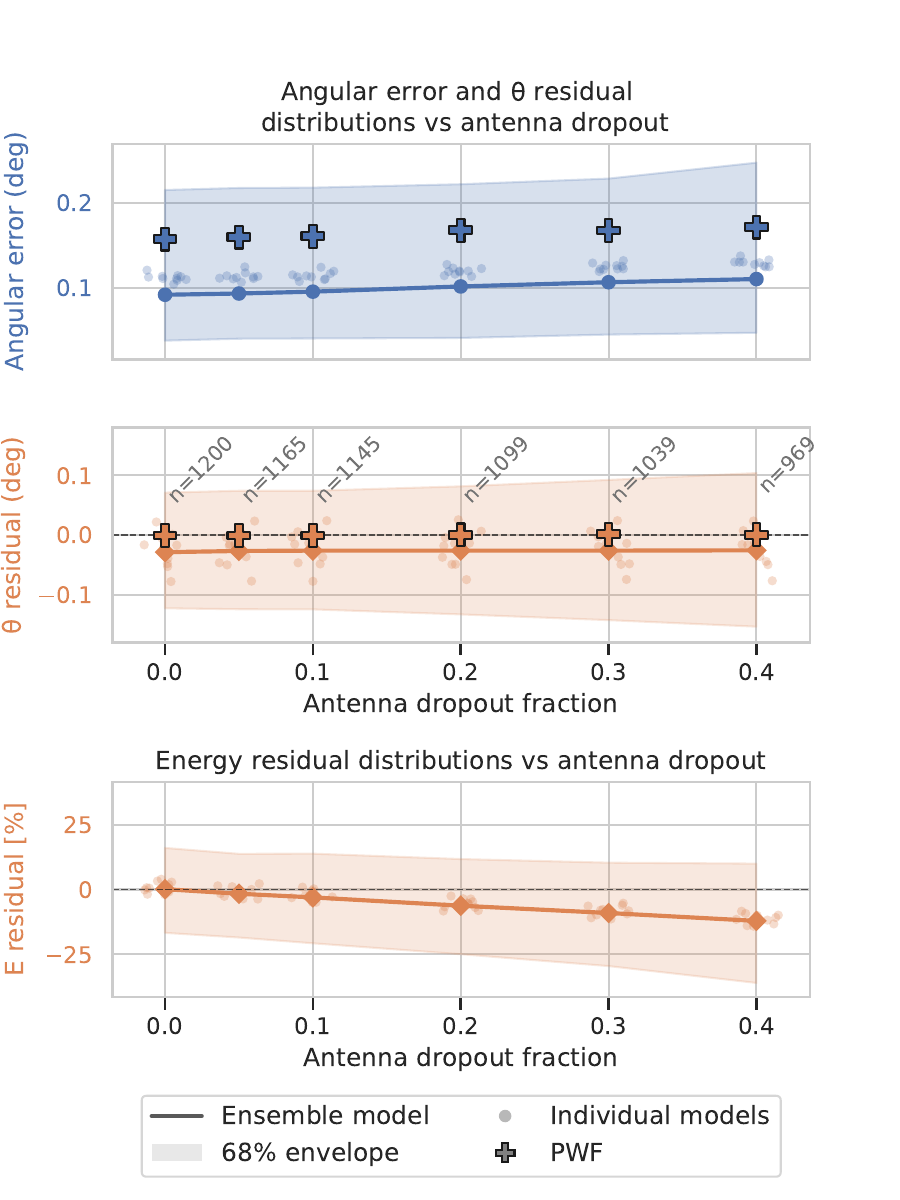}
    \caption{Angular resolution (median values) and zenith angle bias (mean values) as a function of the fraction of deactivated antennas for the ensemble model (solid line) and the PWF estimator (crosses) and single models (transparent dots). The angular resolution degrades from 0.09° to 0.11° with 40\% antennas deactivated. Bottom: Energy resolution as a function of the fraction of deactivated antennas for the ensemble model. The energy resolution worsens from 16.4\% to 23.0\% with 40\% antennas deactivated.}\label{fig:robustness_dropout}
\end{figure}

 As shown in Fig.~\ref{fig:robustness_dropout}, the angular resolution degrades only marginally, increasing from 0.09$^\circ$ to 0.11$^\circ$ at 40\% dropout. No significant bias is observed in the reconstruction of $\theta$ or $\phi$, and the GNN consistently outperforms the PWF baseline across all configurations, indicating a robust correction of the plane-wave approximation under reduced detector sampling. \\

The energy reconstruction is more sensitive to antenna dropout, with the resolution degrading from 16.4\% to 23.0\% in the extreme case of 40\% dropout. A bias in the reconstructed energy also becomes more pronounced as the dropout fraction increases. This behavior is expected, as random antenna removal modifies the spatial sampling of the radio footprint in a way that is not fully represented in the training distribution, rather than simply reducing the signal amplitude.

This results in a distribution shift between training and evaluation conditions, which explains the observed degradation in energy reconstruction performance.

\subsection{Antenna gain variation}
The systematic uncertainties on amplitude calibration are estimated to a value of 7.5\% in GRAND (see Section~\ref{sec:noise}), but imperfect calibration procedure or specific antenna conditions may degrade this value. To evaluate the effect of such antenna gain variations, we multiply for each antenna the RF transfer function $\mathbfcal{R}(\nu)$ introduced in Section~\ref{sec:voltage} by a Gaussian random factor with mean 1 and relative standard deviation of 0\%, 5\%, 10\%, 15\%, and 20\%.

The same set of 1200 validation events are used for all configurations, and the trigger threshold is maintained to the 5$\sigma$ value to isolate the effect of gain variations. 

\begin{figure}
    \centering
    \includegraphics[width=0.65\textwidth]{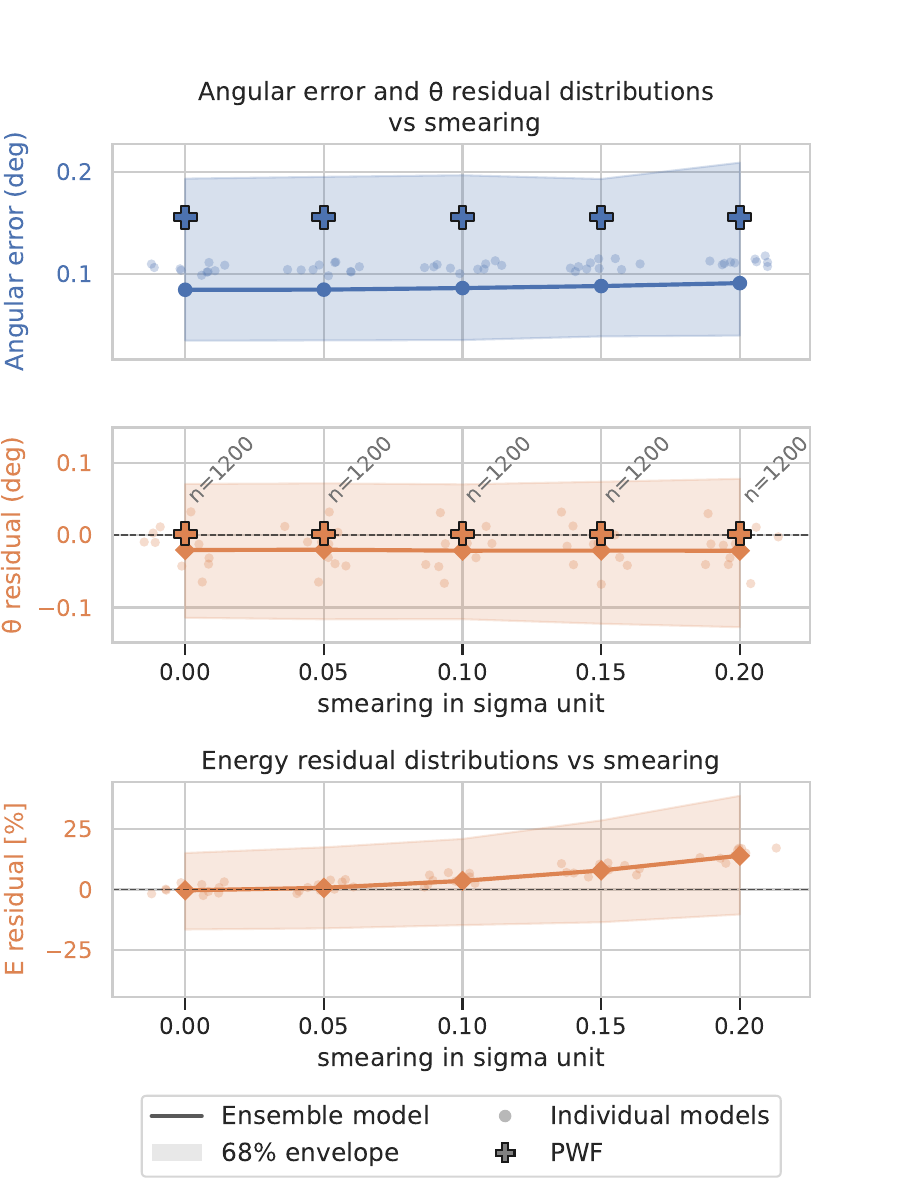}
    \caption{Angular resolution (median values) and zenith angle bias (mean values) as a function of miscalibration for the ensemble model (solid line), the PWF estimator (crosses) and single models (transparent dots). The angular performances don't degrade. Bottom: Energy resolution as a function of the fraction of miscalibration for the ensemble model and single model. The energy resolution worsens from 15.9\% to 24.0\% with 20\% smearing on antenna amplitudes. A bias also appears above 10\% miscalibration}\label{fig:robustness_gain}
\end{figure}

As shown in Fig.~\ref{fig:robustness_gain}, the angular resolution remains stable across all configurations, indicating that the direction reconstruction is only weakly sensitive to moderate gain variations.

In contrast, the energy reconstruction is more sensitive to amplitude perturbations, with the resolution degrading from 16.4\% in the nominal case to 24.4\% at 20\% smearing. A bias is also observed in the reconstructed energy, reaching approximately +13.9\% at the highest smearing level. This behavior is expected, as the energy estimator relies directly on signal amplitudes, which are affected by gain variations, whereas the direction reconstruction is primarily driven by timing information.

\section{Conclusion}\label{section:conclusions}
In this work, we have developed a deep ensemble graph neural network architecture to reconstruct cosmic-ray direction and energy in GRAND-like radio antenna arrays. The model is trained on a simulation dataset designed to be both realistic for current experimental conditions and sufficiently general to remain applicable beyond the specific GRANDProto300 configuration, which served as a reference for the instrumental setup.

By combining array geometry, physics-informed features extracted from voltage signals, and prior information from a plane-wavefront (PWF) reconstruction, the ensemble model achieves excellent direction reconstruction performance, reaching a median angular resolution of 0.092$^\circ$, compared to 0.16$^\circ$ for the PWF method. This performance is comparable to established physics-based approaches applied to electric-field measurements (e.g. ADF) in similar array geometries, while relying solely on voltage information and thus avoiding explicit antenna-response deconvolution.

For energy reconstruction, the model achieves a resolution of 16.4\% on the electromagnetic component using only voltage-based inputs. While direct comparisons with existing voltage-based reconstruction methods remain limited, these results highlight the potential of learning-based approaches in this regime.

In addition, the model provides event-by-event uncertainty estimates that are well calibrated and useful for downstream analyses.

We further evaluate the robustness of the method under variations in experimental conditions, including changes in trigger thresholds, antenna gain fluctuations, and random antenna dropouts. Direction reconstruction remains stable under all tested conditions, whereas energy reconstruction shows a stronger sensitivity, particularly under antenna dropout scenarios, highlighting the importance of representative training datasets that include realistic detector variations.

Overall, these results demonstrate that deep ensemble GNNs are a promising approach for improving cosmic-ray reconstruction in radio detection arrays.The combination of machine learning and domain knowledge provides a powerful pathway toward enhanced radio-based cosmic-ray measurements. Future work will focus on incorporating additional physics-informed features, extending the method to alternative detector geometries, and applying the framework to real observational data. A first test on real data already shows promising agreement with an independent standard analysis.

\section*{Acknowledgements}
We thank the GRAND Collaboration for the discussion and useful insights for the simulations. We also thank the GRAND machine learning group (in particular C. Guépin, O. Macias, S. El Bouch and S. Prunet) for suggestions and discussions around the method. Simulations were performed using the computing resources at the IN2P3 Computing Centre (Villeurbanne, France), a partnership between CNRS/IN2P3 and CEA/DSM/Irfu.

\bibliographystyle{elsarticle-num}
\bibliography{biblio}

@article{Kahn1966,
    author = {Kahn, Franz Daniel and Lerche, I.},
    title = {Radiation from cosmic ray air showers},
    journal = {Proceedings of the Royal Society of London. A. Mathematical and Physical Sciences},
    volume = {289},
    number = {1417},
    pages = {206-213},
    year = {1966},
    month = {01},
    issn = {0080-4630},
    doi = {10.1098/rspa.1966.0007},
    url = {https://doi.org/10.1098/rspa.1966.0007},
    eprint = {https://royalsocietypublishing.org/rspa/article-pdf/289/1417/206/55891/rspa.1966.0007.pdf},
}

@article{GRAND_paper_2019,
	doi = {10.1007/s11433-018-9385-7},
	url = {https://doi.org/10.1007%2Fs11433-018-9385-7},
	year = 2019,
	month = {aug},
	publisher = {Springer Science and Business Media {LLC}},
	volume = {63},
	number = {1},
	author = {{GRAND Collaboration}},
	title = {The Giant Radio Array for Neutrino Detection ({GRAND}): Science and design},
	journal = {Science China Physics, Mechanics and Astronomy}
}

@INPROCEEDINGS{AugerPrime_2019,
       author = {{Castellina}, Antonella and {Pierre Auger Collaboration}},
        title = "{AugerPrime: the Pierre Auger Observatory Upgrade}",
    booktitle = {European Physical Journal Web of Conferences},
         year = 2019,
       series = {European Physical Journal Web of Conferences},
       volume = {210},
        month = oct,
          eid = {06002},
        pages = {06002},
          doi = {10.1051/epjconf/201921006002},
archivePrefix = {arXiv},
       eprint = {1905.04472},
 primaryClass = {astro-ph.HE},
       adsurl = {https://ui.adsabs.harvard.edu/abs/2019EPJWC.21006002C}
}

@article{Ferriere2024,
title = {Analytical planar wavefront reconstruction and error estimates for radio detection of extensive air showers},
journal = {Nucl. Instrum. Methods Phys. Res. A},
volume = {1072},
pages = {170178},
year = {2025},
issn = {0168-9002},
doi = {https://doi.org/10.1016/j.nima.2024.170178},
url = {https://www.sciencedirect.com/science/article/pii/S0168900224011045},
author = {Arsène Ferrière and others}
}

@article{CORSTANJE201522,
title = {The shape of the radio wavefront of extensive air showers as measured with LOFAR},
journal = {Astroparticle Physics},
volume = {61},
pages = {22-31},
year = {2015},
issn = {0927-6505},
doi = {https://doi.org/10.1016/j.astropartphys.2014.06.001},
url = {https://www.sciencedirect.com/science/article/pii/S0927650514000851},
author = {Corstanje, A. and others}
}

@article{GUELFAND2025,
title = {Reconstruction of inclined extensive air showers using radio signals: From arrival times and amplitudes to direction and energy},
journal = {Astroparticle Physics},
volume = {171},
pages = {103120},
year = {2025},
issn = {0927-6505},
doi = {https://doi.org/10.1016/j.astropartphys.2025.103120},
url = {https://www.sciencedirect.com/science/article/pii/S092765052500043X},
author = {Marion Guelfand and others}
}

@article{Edge_conv,
author = {Wang, Yue and others},
title = {Dynamic Graph CNN for Learning on Point Clouds},
year = {2019},
publisher = {Association for Computing Machinery},
address = {New York, NY, USA},
volume = {38},
number = {5},
issn = {0730-0301},
url = {https://doi.org/10.1145/3326362},
doi = {10.1145/3326362},
journal = {ACM Trans. Graph.},
articleno = {146}
}

@article{Garcia_2025,
  author = "Garcia Ginez, Rocio",
  title = "{Graph neural network model to classify between CR and gamma rays in ALPAQUITA (SD)}",
  doi = "10.22323/1.501.0653",
  journal = "PoS",
  year = 2025,
  volume = "ICRC2025",
  pages = "653"
}

@inproceedings{FerriereICRC,
  author = "Ferriere, Arsene  and  others",
  title = "{Reconstruction of cosmic-ray properties with uncertainty estimation using graph neural networks in GRAND}",
  doi = "10.22323/1.501.0253",
  booktitle = "Proceedings of 39th International Cosmic Ray Conference {\textemdash} PoS(ICRC2025)",
  year = 2025,
  volume = "501",
  pages = "253"
}

@article{Koundal_2025,
  author = "Koundal, Paras  and  {ICECUBE Collaboration}",
  title = "{Machine learning driven reconstruction of cosmic-ray air showers for next generation radio arrays}",
  doi = "10.22323/1.501.0309",
  journal = "PoS",
  year = 2025,
  volume = "ICRC2025",
  pages = "309"
}

@article{Guelzow_ICRC,
    author = {G{\"u}lzow, Lukas and Engel, Ralph and Huege, Tim and Roth, Markus and Schl{\"u}ter, Felix},
    title = "{Signal Model and Energy Reconstruction for the Radio Detection of Inclined Air Showers in the 50-200 MHz Frequency Band}",
    eprint = "2507.06698",
    archivePrefix = "arXiv",
    primaryClass = "astro-ph.IM",
    reportNumber = "PoS(ICRC2025)283",
    doi = "10.22323/1.501.0283",
    journal = "PoS",
    volume = "ICRC2025",
    pages = "283",
    year = "2025"
}

@article{Coleman:2022abf,
    author = "Coleman, A. and others",
    title = "{Ultra high energy cosmic rays The intersection of the Cosmic and Energy Frontiers}",
    eprint = "2205.05845",
    archivePrefix = "arXiv",
    primaryClass = "astro-ph.HE",
    reportNumber = "FERMILAB-PUB-22-413-PPD",
    doi = "10.1016/j.astropartphys.2023.102819",
    journal = "Astropart. Phys.",
    volume = "149",
    pages = "102819",
    year = "2023"
}

@article{IceCube_CNN,
    author = "Abbasi, R. and others",
    title = "{A Convolutional Neural Network based Cascade Reconstruction for the IceCube Neutrino Observatory}",
    eprint = "2101.11589",
    archivePrefix = "arXiv",
    primaryClass = "hep-ex",
    doi = "10.1088/1748-0221/16/07/P07041",
    journal = "JINST",
    volume = "16",
    pages = "P07041",
    year = "2021"
}

@ARTICLE{XuXing,
  author={Zhang, Pengfei and Xu, Xin and Wang, Hanrui and Xu, Xing and Duan, Bohao and Wei, Feng and Pan, Hongwei and Tian, Xishui and Zhang, Yi and Ma, Pengxiong and Martineau-Huynh, Olivier},
  journal={IEEE Transactions on Instrumentation and Measurement}, 
  title={Noise Suppression and Radio Frequency Interference Rejection for Self-Triggered Radio Detectors of Extensive Air Showers}, 
  year={2026},
  volume={75},
  number={},
  pages={1-19}}

@article{AERA_calib,
author = {Collaboration, The and Aab, Alexander and Abreu, Pedro and Aglietta, Marco and Al Samarai, Imen and Albuquerque, Ivone and Allekotte, I. and Almela, Alejandro and Castillo, Jesus and Alvarez-Muñiz, Jaime and Anastasi, G.A. and Anchordoqui, Luis and Andrada, Belén and Andringa, Sofia and Aramo, C. and Arqueros, Fernando and Arsene, Nicusor and Asorey, Hernán and Assis, Pedro and Zuccarello, Francesca},
year = {2017},
month = {02},
pages = {},
title = {Calibration of the Logarithmic-Periodic Dipole Antenna (LPDA) Radio Stations at the Pierre Auger Observatory using an Octocopter},
volume = {12},
journal = {Journal of Instrumentation},
doi = {10.1088/1748-0221/12/10/T10005}
}

@book{balanis01,
  author = {Balanis, Constantine A.},
  edition = {Second},
  publisher = {John Wiley Sons},
  title = {Antenna Theory: Analysis and Design},
  username = {dalbem},
  year = 1997
}

@article{NUTRIG,
    author = {Correa, Pablo and K{\"o}hler, Jelena},
    collaboration = "GRAND",
    title = "{NUTRIG: Development of a Novel Radio Self-Trigger for GRAND}",
    eprint = "2507.04339",
    archivePrefix = "arXiv",
    primaryClass = "astro-ph.IM",
    doi = "10.22323/1.501.0229",
    journal = "PoS",
    volume = "ICRC2025",
    pages = "229",
    year = "2025"
}

@article{GRANDHW,
    author = "{\'A}lvarez-Mu{\~n}iz, J. and others",
    collaboration = "GRAND",
    title = "{Towards the Giant Radio Array for Neutrino Detection (GRAND): the GRANDProto300 and GRAND@Auger prototypes}",
    eprint = "2509.21306",
    archivePrefix = "arXiv",
    primaryClass = "astro-ph.IM",
    doi = "10.1088/1748-0221/21/02/P02041",
    journal = "JINST",
    volume = "21",
    number = "02",
    pages = "P02041",
    year = "2026"
}

@article{LDF_Auger,
    author = {Schl{\"u}ter, Felix and Huege, Tim},
    title = "{Signal model and event reconstruction for the radio detection of inclined air showers}",
    eprint = "2203.04364",
    archivePrefix = "arXiv",
    primaryClass = "astro-ph.HE",
    doi = "10.1088/1475-7516/2023/01/008",
    journal = "JCAP",
    volume = "01",
    pages = "008",
    year = "2023"
}

@inproceedings{GRAND_ICRC2025,
  author    = {Martineau-Huynh, Olivier},
  title     = {The Giant Radio Array for Neutrino Detection - experimental status and plans},
  booktitle = {Proceedings of the 39th International Cosmic Ray Conference --- PoS(ICRC2025)},
  volume    = {501},
  pages     = {1114},
  publisher = {SISSA Medialab},
  year      = {2025},
  doi       = {10.22323/1.501.1114}
}

@misc{Macias2025Direction,
    author = "Macias, Oscar and others",
    title = "{Simulation-Based Inference for Direction Reconstruction of Ultra-High-Energy Cosmic Rays with Radio Arrays}",
    eprint = "2508.15991",
    archivePrefix = "arXiv",
    primaryClass = "astro-ph.HE",
    url={https://arxiv.org/abs/2508.15991},
    month = "8",
    year = "2025"
}

@article{Alvarez_Muniz_2012,
   title={Monte Carlo simulations of radio pulses in atmospheric showers using ZHAireS},
   volume={35},
   ISSN={0927-6505},
   url={http://dx.doi.org/10.1016/j.astropartphys.2011.10.005},
   DOI={10.1016/j.astropartphys.2011.10.005},
   number={6},
   journal={Astroparticle Physics},
   publisher={Elsevier BV},
   author={Alvarez-Muñiz, Jaime and others},
   year={2012},
   month=jan, pages={325–341} }

@article{layout2024,
author = {Benoit-Lévy, Aurélien and others},
year = {2024},
month = {04},
pages = {P04006},
title = {Pruning: a tool to optimize the layout of large scale arrays for ultra-high-energy air-shower detection},
volume = {19},
journal = {Journal of Instrumentation},
doi = {10.1088/1748-0221/19/04/P04006}
}

@article{GRANDLIB_2025,
    title = {{GRANDlib: A simulation pipeline for the Giant Radio Array for Neutrino Detection (GRAND)}},
    journal = {Computer Physics Communications},
    volume = {308},
    pages = {109461},
    year = {2025},
    issn = {0010-4655},
    doi = {https://doi.org/10.1016/j.cpc.2024.109461},
    url = {https://www.sciencedirect.com/science/article/pii/S0010465524003849},
    author = {{GRAND Collaboration}}
    }

@misc{GRANDProtoPapers_2025,
      title={Towards the Giant Radio Array for Neutrino Detection ({GRAND}): the {GRANDProto300} and {GRAND@Auger} prototypes}, 
      author = {{GRAND Collaboration}},
      year={2025},
      eprint={2509.21306},
      archivePrefix={arXiv},
      primaryClass={astro-ph.IM},
      url={https://arxiv.org/abs/2509.21306},
      note={Accepted for pub. in JINST}
}

@article{Decoene_2023,
   title={Radio wavefront of very inclined extensive air-showers: A simulation study for extended and sparse radio arrays},
   volume={145},
   ISSN={0927-6505},
   url={http://dx.doi.org/10.1016/j.astropartphys.2022.102779},
   DOI={10.1016/j.astropartphys.2022.102779},
   journal={Astroparticle Physics},
   publisher={Elsevier BV},
   author={Decoene, Valentin and others},
   year={2023},
   month=mar, pages={102779} }

@article{polisensky_2007,
  title={LFmap: A low frequency sky map generating program},
  author={Polisensky, Emil},
  journal={Long Wavelength Array Memo Series},
  volume={111},
  pages={515},
  year={2007}
}

@INPROCEEDINGS{NLL_1994,
  author={Nix, D.A. and Weigend, A.S.},
  booktitle={Proceedings of 1994 IEEE International Conference on Neural Networks (ICNN'94)}, 
  title={Estimating the mean and variance of the target probability distribution}, 
  year={1994},
  volume={1},
  number={},
  pages={55-60 vol.1},
  doi={10.1109/ICNN.1994.374138}}

@article{chiche_decoherence,
  title = {Loss of Coherence and Change in Emission Physics for Radio Emission from Very Inclined Cosmic-Ray Air Showers},
  author = {Chiche, Simon and others},
  journal = {Phys. Rev. Lett.},
  volume = {132},
  issue = {23},
  pages = {231001},
  numpages = {6},
  year = {2024},
  month = {Jun},
  publisher = {American Physical Society},
  doi = {10.1103/PhysRevLett.132.231001},
  url = {https://link.aps.org/doi/10.1103/PhysRevLett.132.231001}
}

@ARTICLE{Auger_gal,
       author = {Halim, Abdul and others},
        title = "{Long-term calibration and validation of stability of the Auger Engineering Radio Array using the diffuse Galactic radio emission}",
      journal = {Journal of Instrumentation},
         year = 2025,
        month = dec,
       volume = {20},
       number = {12},
          eid = {P12017},
        pages = {P12017},
          doi = {10.1088/1748-0221/20/12/P12017},
archivePrefix = {arXiv},
       eprint = {2512.03692},
 primaryClass = {astro-ph.IM},
       adsurl = {https://ui.adsabs.harvard.edu/abs/2025JInst..20P2017A}
}

@article{Charrier_2019,
   title={Autonomous radio detection of air showers with the TREND50 antenna array},
   volume={110},
   ISSN={0927-6505},
   url={http://dx.doi.org/10.1016/j.astropartphys.2019.03.002},
   DOI={10.1016/j.astropartphys.2019.03.002},
   journal={Astroparticle Physics},
   publisher={Elsevier BV},
   author={Charrier, D. and others},
   year={2019},
   month=jul, pages={15–29} }

@inproceedings{Burke2004NEC,
  author={Burke, G.J. and others},
  booktitle={IEEE Antennas and Propagation Society Symposium, 2004.}, 
  title={The Numerical Electromagnetics Code (NEC) - a brief history}, 
  year={2004},
  volume={3},
  number={},
  pages={2871-2874 Vol.3},
  doi={10.1109/APS.2004.1331976}}

\appendix
\section{Influence of the training size on the reconstruction performances}\label{app:training_size}
For the training of the GNN models, we use up to 5000 realistic simulated events. In this appendix, we study the dependence of the reconstruction performance on the size of the training dataset. To this end, we train models using 300, 500, 1000, 2000, 3000, 4000, and 5000 training events. For each dataset size, we train 10 independent models and evaluate the performance of both the individual models and their ensemble. This procedure is performed for the models described in this paper as well as for models where no prior information from the PWF are given. The former will be called here pGNN (for PWF-GNN) and the latter rGNN (for raw-GNN). All other training hyper-parameters and model architectures are kept identical across dataset sizes and architectures. The validation set used is the same across the whole experiment.

Figures~\ref{fig:energy_dataset_size} and~\ref{fig:angular_dataset_size} show the evolution of the energy resolution and angular resolution, respectively, as defined in Section~\ref{section:performances}.

For the energy reconstruction, we observe an approximately linear relation between the resolution and the log of the training set size, suggesting further performance gains for larger training datasets. We also find that the inclusion of the PWF-based features does not significantly affect the energy reconstruction performance. 

For direction reconstruction, models incorporating PWF-based inputs show a clear improvement over the rGNN models. We further observe that the performance of the pGNN models stops improving for training set sizes above approximately 2000 events, whereas the rGNN models continue to improve up to 5000 training events. This behavior supports the use of PWF cartesian coordinates as inputs to GNNs for direction reconstruction when the size of the training set is limited. The reported trends are robust across different random initializations, as quantified by the spread among the individual models in each ensemble.

\begin{figure}
    \centering
    \includegraphics[width=0.6\textwidth]{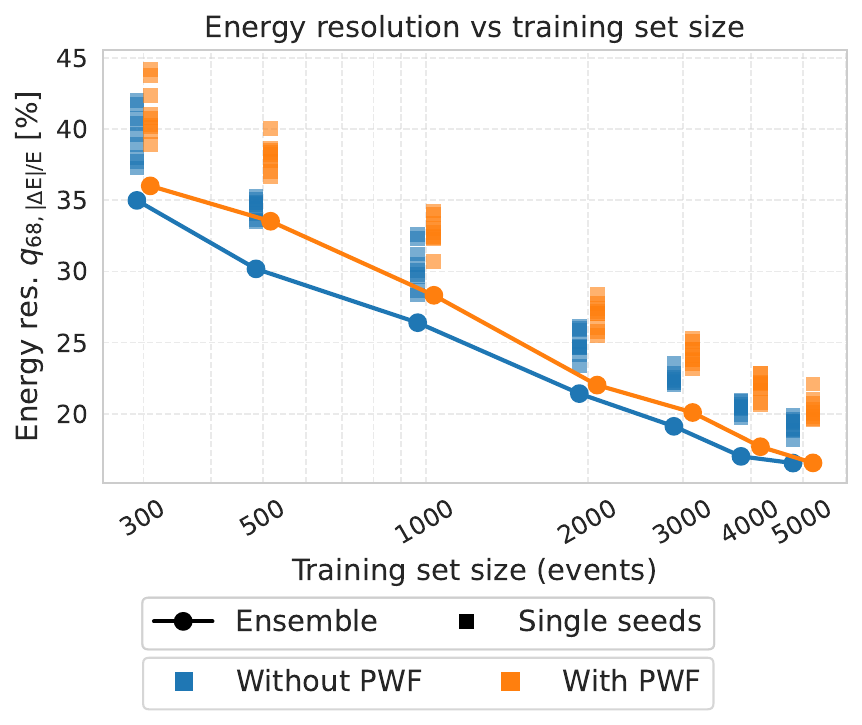}
    \caption{Energy resolution as a function of training set size. Ensemble predictions (circles) and individual seed predictions (squares) are shown for models trained with (orange) and without (blue) PWF input. Energy resolution is defined as the 68\% quantile on energy relative error $q_{68, |\Delta \rm{E}| / \rm{E}}$.}\label{fig:energy_dataset_size}
\end{figure}

\begin{figure}
    \centering
    \includegraphics[width=0.6\textwidth]{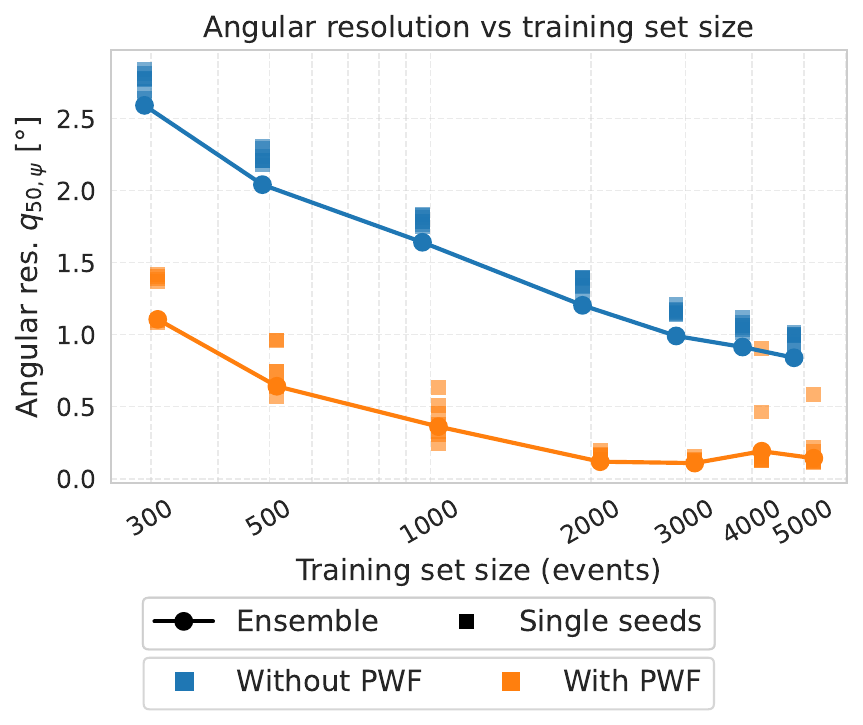}
    \caption{Angular resolution as a function of training set size for direction reconstruction. Ensemble predictions (circles) and individual seed predictions (squares) are shown for models trained with (orange) and without (blue) point spread function (PWF) corrections. Angular resolution is defined as the median angular error $q_{50,\psi}$ in degrees.}\label{fig:angular_dataset_size}
\end{figure}

\section{Propagation of uncertainties for direction reconstruction}\label{app:propa_theta_phi}
In this section, we will derive how an uncertainty on the direction vector $\rbm{k}$ can be propagated to its polar coordinates.
The direction models predict the Cartesian components of the unit vector $\hat{\rbm{k}}$ together with per-component standard deviations $\hat{\sigma}_x, \hat{\sigma}_y, \hat{\sigma}_z$. We form the (diagonal) covariance
\begin{equation}
 \hat{\Sigma}_{ij} = \hat{\sigma}_i^2\,\delta_{ij},
\qquad i,j \in \{x,y,z\}.
\end{equation}
The underlying assumption is that the the uncertainty on $\rbm{k}$ is Gaussian with mean $\hat{\rbm{k}}$ and covariance matrix $\hat{\Sigma}$.
Because the physical direction lies on the unit sphere ($\rbm{k}$ is a unit vector), we project this unconstrained Gaussian onto the tangent plane at $\hat{\rbm{k}}$. Using the Gaussian conditioning formula for the linear constraint $\hat{\rbm{k}}^{\!T}(\rbm{k}-\hat{\rbm{k}})=0$, the constrained covariance is
\begin{equation}\label{eq:DistNoConstraint}
    \hat{\bm{\Sigma}}_{\perp} \;=\;
    \hat{\bm{\Sigma}} \;-\;
    \frac{\hat{\bm{\Sigma}}\,\hat{\rbm{k}}\,\hat{\rbm{k}}^{\!T}\,\hat{\bm{\Sigma}}}{\hat{\rbm{k}}^{\!T}\,\hat{\bm{\Sigma}}\,\hat{\rbm{k}}}\,.
\end{equation}
We then propagate uncertainties to the spherical angles $(\uptheta,\upphi)$. Two equivalent routes can be used.
\begin{enumerate}
    \item Direct Jacobian from Cartesian to angles (valid away from the poles, $\sin\uptheta \neq 0$):
    \begin{align}
        J &= \begin{pmatrix}
        \dfrac{\partial \uptheta}{\partial k_x} & \dfrac{\partial \uptheta}{\partial k_y} & \dfrac{\partial \uptheta}{\partial k_z} \\
        \dfrac{\partial \upphi}{\partial k_x}   & \dfrac{\partial \upphi}{\partial k_y}   & \dfrac{\partial \upphi}{\partial k_z}
        \end{pmatrix}\\
        &=
        \begin{pmatrix}
            0 & 0 & \dfrac{1}{\sin\uptheta} \\
            \dfrac{\sin\upphi}{\sin\uptheta} & -\dfrac{\cos\upphi}{\sin\uptheta} & 0
        \end{pmatrix},\\
        \hat{\bm{\Sigma}}_{\uptheta,\upphi} &= J\,\hat{\bm{\Sigma}}_{\perp}\,J^{\!T}.
    \end{align}
    \item Using the Schur complement inversion method, the previous expression is equivalent to the following, which is numerically more robust near the poles:
    
    \begin{equation}
        \hat{\bm{\Sigma}}_{\uptheta,\upphi} \;=\;
        \Big{[\, \rbm{R_a}^{\!\rm T}\,\hat{\bm{\Sigma}}^{-1}\,\rbm{R_a}\,\Big]}^{-1},\
    \end{equation}
    with the Jacobian from angles to Cartesian (using Eq.~\ref{eq:direction_vector}):
    \begin{equation}
        \rbm{R_a} \;=\;
        \begin{pmatrix}
            \dfrac{\partial k_x}{\partial \uptheta} & \dfrac{\partial k_x}{\partial \upphi} \\
            \dfrac{\partial k_y}{\partial \uptheta} & \dfrac{\partial k_y}{\partial \upphi} \\
            \dfrac{\partial k_z}{\partial \uptheta} & \dfrac{\partial k_z}{\partial \upphi}
        \end{pmatrix}
        =
        \begin{pmatrix}
            -\cos\uptheta\cos\upphi & \;\;\sin\uptheta\sin\upphi \\
            -\cos\uptheta\sin\upphi & -\sin\uptheta\cos\upphi \\
            \;\;\sin\uptheta        & \;\;0
    \end{pmatrix}.
    \end{equation}
\end{enumerate}

\end{document}